\documentclass[12pt,preprint]{aastex}

\newcommand{\myemail}{bsiana@ipac.caltech.edu}
\newcommand{\leqsim}{\,\raisebox{-0.6ex}{$\buildrel < \over \sim$}\,}
\newcommand{\geqsim}{\,\raisebox{-0.6ex}{$\buildrel > \over \sim$}\,}

\shorttitle{Escape Fraction at $z \sim 1.3$.}
\shortauthors{Siana et al.}

\begin{document}


\title{New Constraints on the Lyman Continuum Escape Fraction at $z \sim 1.3^{\dagger}$}


\author{Brian Siana, Harry I. Teplitz, James Colbert\altaffilmark{1}, Henry C. Ferguson\altaffilmark{2}, Mark Dickinson \altaffilmark{3}, Thomas M. Brown\altaffilmark{2}, Christopher J. Conselice \altaffilmark{4}, Duilia F. de Mello\altaffilmark{5,6}, Jonathan P. Gardner\altaffilmark{6}, Mauro Giavalisco \altaffilmark{7}, Felipe Menanteau\altaffilmark{8}}

\email{\myemail}

\altaffiltext{1}{$Spitzer$ Science Center, California Institute of Technology, 220-6, Pasadena, CA 91125}

\altaffiltext{2}{Space Telescope Science Institute, 3700 San Martin Drive, Baltimore, MD 21218}

\altaffiltext{3}{National Optical Astronomy Observatory, 950 N. Cherry Ave., Tucson, AZ 85719}

\altaffiltext{4}{University of Nottingham, Nottingham, NG7 2RD, UK}

\altaffiltext{5}{Department of Physics, Catholic University of America, 620 Michigan Avenue, Washington DC 20064}

\altaffiltext{6}{Astrophysics Science Division, Observational Cosmology Laboratory, Code 665, Goddard Space Flight Center, Greenbelt, MD 20771}

\altaffiltext{7}{University of Massachusetts, Department of Astronomy, Amherst, MA, 01003}

\altaffiltext{8}{Department of Physics and Astronomy, Rutgers University, 136 Frelinghuysen Road, Piscataway, NJ 08854}

\altaffiltext{${\dagger}$}{Based on observations made with the NASA/ESA Hubble Space Telescope, obtained at the Space Telescope Science Institute, which is operated by the Association of Universities for Research in Astronomy, Inc., under NASA contract NAS 5-26555.  These observations are associated with programs 7410, 9478, and 10403 (as well as 6337 HDF-WFPC2, 7817 HDF-NICMOS, 9978,10086 HUDF-ACS,9803 HUDF-NICMOS).}

\begin{abstract}

We examine deep far-ultraviolet (1600\AA) imaging of the Hubble Deep Field-North (HDFN) and the Hubble Ultra Deep Field (HUDF) to search for leaking Lyman continuum radiation from starburst galaxies at $z \sim 1.3$.  There are 21 (primarily sub-$L^*$) galaxies with spectroscopic redshifts between $1.1<z<1.5$ and none are detected in the far-UV.  We use two methods to estimate limits to the escape fraction ($f_{esc}$) for these galaxies.  First, to compare with previous works, we assume a fixed 1500 \AA\ to Lyman continuum ratio, $f_{1500}/f_{700}$, intrinsic to the stellar SEDs to convert our $f_{700}$ limits to relative escape fractions.  Second, we fit stellar population templates to the galaxies' optical/near-infrared SEDs to determine the starburst age and level of dust attenuation for each individual galaxy, allowing a more accurate estimate of the intrinsic $f_{1500}/f_{700}$.  We find that the two techniques do not differ significantly if the galaxies are currently undergoing star-formation.  We show that previous high-redshift studies may have underestimated the amplitude of the Lyman Break, and thus the relative escape fraction, by a factor $\sim2$.  Once the starburst age and intergalactic HI absorption are accounted for, 18 galaxies in our sample have limits to the relative escape fraction, $f_{esc,rel}<1.0$ with some limits as low as $f_{esc,rel}<0.10$ and a stacked limit of $f_{esc,rel}<0.08$.  This demonstrates, for the first time, that most sub-$L^*$ galaxies at high redshift do not have large escape fractions.  When combined with a similar study of more luminous galaxies at the same redshift we show that, if all star-forming galaxies at $z \sim 1$ have similar relative escape fractions, the value must be less than 0.14 ($3\sigma$).  We also show that less than 20\% ($3\sigma$) of star-forming galaxies at $z \sim 1$ have relative escape fractions near unity.   These limits contrast with the large escape fractions found at $z\sim3$ and suggest that the average escape fraction has decreased between $z \sim 3$ and $z \sim 1$.

\end{abstract}



\section{Introduction}
\label{introduction}
Both active galactic nuclei (AGN) and young massive stars produce Lyman continuum radiation capable of ionizing hydrogen in the surrounding Intergalactic Medium (IGM).  The fraction of this photoionizing continuum which escapes into the IGM, hereafter referred to as the escape fraction ($f_{esc}$), is a key parameter in determining how both star-formation and black hole growth drive the evolution of the ionization state of the IGM.  In particular, an accurate assesment of the average $f_{esc}$ will help determine if star-formation, rather than active galactic nuclei (AGN), is responsible for hydrogen reionization at $z>6$. 

For the most luminous, unobscured AGN (QSOs), the majority of the Lyman continuum escapes into the IGM \citep{telfer02}.  Since QSO luminosity functions and average spectral energy distributions (SEDs) are well known, their contribution to the intergalactic photoionizing continuum can be accurately estimated.  However, the contribution to the ionizing background from star-formation is far more difficult to determine because large reservoirs of HI and dust within the galaxy absorb the bulk of the Lyman continuum.  

Studies of the mean transmission through the Ly$\alpha$ forest have measured the total HI-ionizing background in the IGM as a function of redshift \citep{mcdonald01,scott02,bolton05,fan06}.  Recent measurements of the QSO luminosity functions at $z < 2$ \citep{richards05} indicate that QSOs alone provide the vast majority of the intergalactic HI ionizing flux at these redshifts \citep{inoue06}.  At $z \geq 2.5$, the QSO space density falls off rapidly and recent measurements of the QSO luminosity functions at $z \sim 3$ indicate that star-formation may contribute at least as much to the ionizing background as QSOs \citep{hunt04,siana07a,fontanot07}.  Since the QSO space density falls off faster than the star-formation density, it is inferred that young massive stars must provide the ionizing flux responsible for reionizing the universe by $z \sim 6$ \citep{inoue06}.  

The first attempts to measure $f_{esc}$ examined low-redshift galaxies with space based ultraviolet (UV) telescopes.  \citet{leitherer95} took UV spectra of four local starbursts which, after accounting for absorption by HI in the Milky Way \citep{hurwitz97}, give upper limits of $f_{esc}<0.032-0.57$.  \citet{deharveng01} took a deep FUSE spectrum of Mrk 45 and derived an upper limit of $f_{esc}<0.06$.  

At higher redshift, \citet[hereafter M03]{malkan03} took deep HST/STIS broadband imaging ($\lambda \sim 1600$\AA) of the Lyman continuum of 11 starbursts at $z \sim 1.2$ and derived upper limits to the escape fraction (here, $f_{esc,rel}$, defined relative to the dust-attenuated continuum flux at 1500 \AA; see Section \ref{esc_frac}) $f_{esc,rel} \leqsim 0.10-0.40$.  \citet[hereafter S01]{steidel01} produced a composite spectrum of 29 Lyman Break Galaxies (LBGs) at $z \sim 3.4$ and detected significant signal below the Lyman continuum.  The measured $L(1500\AA)/L(900\AA) = 4.6 \pm 1.0$ implies that there is little to no absorption of the Lyman continuum beyond what is seen in the UV ($lambda_{rest} \sim 1500$\AA), indicating very small column densities of HI along the line-of-sight through the ISM of these LBGs.  The authors note that, because of a selection bias, this sample is among the bluest quartile of the LBG population and is therefore younger and/or less dust-reddened than the average LBG.  However, other measurements at $z\sim3$ using deep optical spectroscopy \citep{giallongo02}, narrow-band imaging \citep{inoue05}, and broad-band imaging \citep{fernandez-soto03}, have all resulted in upper limits for $f_{esc,rel}$ between 0.04 and 0.72.

The first detection of escaping Lyman continuum from an individual galaxy was reported by \citet{bergvall06} with a 16ks FUSE spectrum of Haro11, giving an escape fraction $0.04<f_{esc}<0.1$, or a relative escape fraction of $0.17<f_{esc,rel}<0.42$ (though see Grimes et al. (2007, submitted) who see no detection down to $f_{esc}<0.02$ with a re-reduction of the same data).  Haro11 is a luminous, metal poor, blue compact galaxy - attributes which are similar to LBGs at $z\sim3$.  However, it is somewhat underluminous ($L_{UV} = 0.2 L^*_{LBG}$), may have a significantly larger halo mass than typical high-redshift LBGs, and was preselected as having a very low HI gas mass $M(HI) \sim 10^8 M_{\odot}$ \citep[unpublished result noted in ][]{bergvall06}.  

Most recently, \citet[hereafter S06]{shapley06} published deep optical spectra of an additional 14 LBGs with an average $f_{esc,rel} \sim 0.14$, about 4.5 times lower than the \citet{steidel01} result.  Two of the galaxies appear to have $f_{esc,rel}\sim1$, while the remaining twelve are undetected.  This suggests that 10-20\% of LBGs have large escape fractions, or that the covering fraction of neutral HI and/or dust is 80-90\% in these systems.  This may be the case if mergers or supernovae chimneys \citep{fujita03} sufficiently disturb the HI reservoirs, producing low-column density lines-of-sight (LOSs) through the galaxies.  

With only three individual detections of escaping Lyman continuum, it is difficult to determine whether any physical parameters (starburst age, luminosity, dust, HI mass, morphology) correlate with $f_{esc}$.  For example, the two detections from \citet{shapley06} do not stand out from the rest of the sample and are, in fact, very different from one another.  Furthermore, most programs to date have necessarily targeted the more luminous starbursts, so the faint end of the luminosity function has not been sufficiently examined. 

We use deep far-UV imaging of the Hubble Deep Field North \citep[HDFN,][]{williams96} and the Hubble Ultra Deep Field \citep[HUDF,][]{beckwith06} to search for the escaping ionizing flux from moderate redshift galaxies.  The observations are taken with the f25QTZ filter on the Space Telescope Imaging Spectrograph (STIS) and the F150LP filter on the Solar-Blind Channel (SBC) of the Advanced Camera for Surveys (ACS) aboard the Hubble Space Telescope (HST).  The F25QTZ and F150LP broadband filters have effective wavelengths of $\sim$1600\AA, providing limits on the Lyman continuum luminosities of 24 star-forming galaxies at $z \sim 1.3$.  These galaxies are typically less luminous than those of previous works \citep{steidel01,malkan03,shapley06}, allowing us to probe trends with luminosity.  In Section \ref{esc_frac} we examine the various assumptions made about galaxy properties which affect our derivation of the escape fraction.  Section \ref{observations} explains the data set and sample selection.  In Section \ref{ratios} we use a simple method of applying these assumed corrections to the $f_{1500}/f_{700}$ flux ratio to estimate the escape fraction.  In Section \ref{sed}, we fit stellar population models to the galaxies' SEDs to determine the starburst age and dust attenuation.  In Section \ref{discussion}, we compare the differences of these two methods and compare our limits to other surveys.  

Throughout the text, we use a flat $\Lambda$CDM Cosmology with $H_0=70$ km s$^{-1}$ Mpc$^{-1}$, $\Omega_m = 0.3$, and $\Omega_{\Lambda}=0.7$.  All flux ratios use units of flux per unit frequency.  

\section{Determining the Escape Fraction}
\label{esc_frac}

There are two commonly used definitions of the escape fraction in the literature.  The {\it absolute} escape fraction, $f_{esc}$, is simply the fraction of emitted Lyman continuum photons (typically measured at 900\AA) that escapes into the IGM.  This definition is useful in theoretical or semi-analytic models where it is used to directly translate star-formation rates and initial mass functions (IMFs) into ionizing backgrounds.  Observationally, the absolute escape fraction is difficult to determine because the intrinsic Lyman continuum flux must first be known.  At low redshift, the H$\alpha$ line flux (corrected for reddening by measuring the Balmer decrement) is often used to approximate the number of Lyman continuum photons emitted from massive stars.  However, at high redshift this becomes increasingly difficult as the Balmer lines shift to the near-IR.  Therefore, the Lyman continuum flux is often inferred from the galaxy's rest-frame UV and optical photometry, in particular the rest-frame 1500\AA\ flux, $f_{\nu}(1500$\AA$)$, since that is easily measured at $z>1$.  Unfortunately, the 1500\AA\ flux is subject to significant dust attenuation so this must be accurately accounted for in order to use it for determining the intrinsic Lyman continuum flux.

\citet{steidel01} defined the {\it relative} escape fraction, $f_{esc,rel}$ as the fraction of escaping Lyman-continuum photons divided by the fraction of escaping photons at 1500\AA.  This definition doesn't require knowledge of the level of dust attenuation, and is useful at high redshift because $f_{1500}$ is easily measured and is commonly used when determining luminosity functions.  Therefore, the relative escape fraction can be used to directly convert luminosity functions to ionizing backgrounds.

In practice, $f_{esc,rel}$ is determined by comparing the Lyman continuum flux with the rest-frame 1500\AA\ flux, $f_{1500}$.  Spectroscopic and narrow-band studies are sensitive to Lyman continuum flux at wavelengths just short of the Lyman Break ($\sim900$\AA), whereas broadband photometric observations probe the Lyman continuum at somewhat shorter wavelengths (typically $\sim 700$ \AA).  This observed flux ratio between 1500\AA\ and the Lyman continuum is affected by several factors and can be expressed as the following product

\begin{equation}
\label{eqn: flux_ratio}
(f_{1500}/f_{LC})_{obs} = (f_{1500}/f_{LC})_{stel} \times 10^{-0.4(A_{1500}-A_{LC})} \times \textrm{exp}(\tau_{HI-IGM}(LC)) \times \textrm{exp}(\tau_{HI-ISM}(LC)) 
\end{equation}

{\noindent}where LC is the wavelength at which the Lyman continuum is being observed ($\sim700$\AA\ in this study),  $(f_{1500}/f_{LC})_{stel}$ is the intrinsic flux ratio from the SED of the stellar population,  $(A_{1500}-A_{LC})$ is the differential dust attenuation (in magnitudes), $\tau_{HI-IGM}(LC)$ is the optical depth of the Lyman line and continuum absorption through the IGM along the line-of-sight to that galaxy, and $\tau_{HI-ISM}(LC)$ is the optical depth of the Lyman continuum absorption from HI within the observed galaxy's interstellar medium (ISM).  

We can rearrange this equation to give us the {\it relative} escape fraction

\begin{equation}
\label{eqn: fesc_rel}
f_{esc,rel} = \frac{(f_{1500}/f_{LC})_{stel}}{(f_{1500}/f_{LC})_{obs}} \times \textrm{exp}(\tau_{HI-IGM}(LC)) = \textrm{exp}(-\tau_{HI-ISM}(LC)) \times 10^{0.4(A_{1500}-A_{LC})} .
\end{equation} 

{\noindent}If we can estimate the amplitude of the intrinsic stellar Lyman break, $(f_{1500}/f_{LC})_{stel}$, and the average optical depth of the Ly$\alpha$ forest, $\tau_{HI-IGM}(LC)$, then then relative escape fraction can be computed directly from the observed flux ratio, $(f_{1500}/f_{LC})_{obs}$.  In this section we detail each of these factors and quantify their effects on the $(f_{1500}/f_{LC})_{obs}$ ratio.  Note that the relative escape fraction is the product of the HI absorption and the {\it differential} reddening between 1500\AA\ and the Lyman continuum, whereas the absolute escape fraction is the product of the HI absorption and the {\it total} dust attenuation of the Lyman continuum.  Therefore, the absolute escape fraction is equal to the relative escape fraction times the dust attenuation at 1500 \AA,

\begin{equation}
\label{eqn: fesc_conv}
f_{esc} = 10^{-0.4A(1500)} f_{esc,rel}
\end{equation}

{\noindent}where $A(1500) = 10.33 E(B-V)$ for a Calzetti reddening law \citep{calzetti97}.

\subsection{Stellar Population}
\label{stellar}
The intrinsic flux decrement across the Lyman Break is dependent upon the age, star-formation history (single burst, exponential decay, constant), initial mass-function (IMF), and metallicity of the stellar population.  We have used both the Starburst99 \citep{leitherer99} and BC03 \citep{bruzual03} population synthesis models to quantify how these properties affect the size of the Lyman Break in star-forming galaxies.  We note that the non-LTE models which better account for stellar winds and line blanketing in the atmospheres of massive stars \citep{schaerer98,smith02} do not significantly affect the HI ionizing continuum between $700<\lambda<900$\AA\ \citep[see Figure 8 in][]{smith02}.

\subsubsection{Star Formation History}

\begin{figure}
\epsscale{1.0}
\plotone{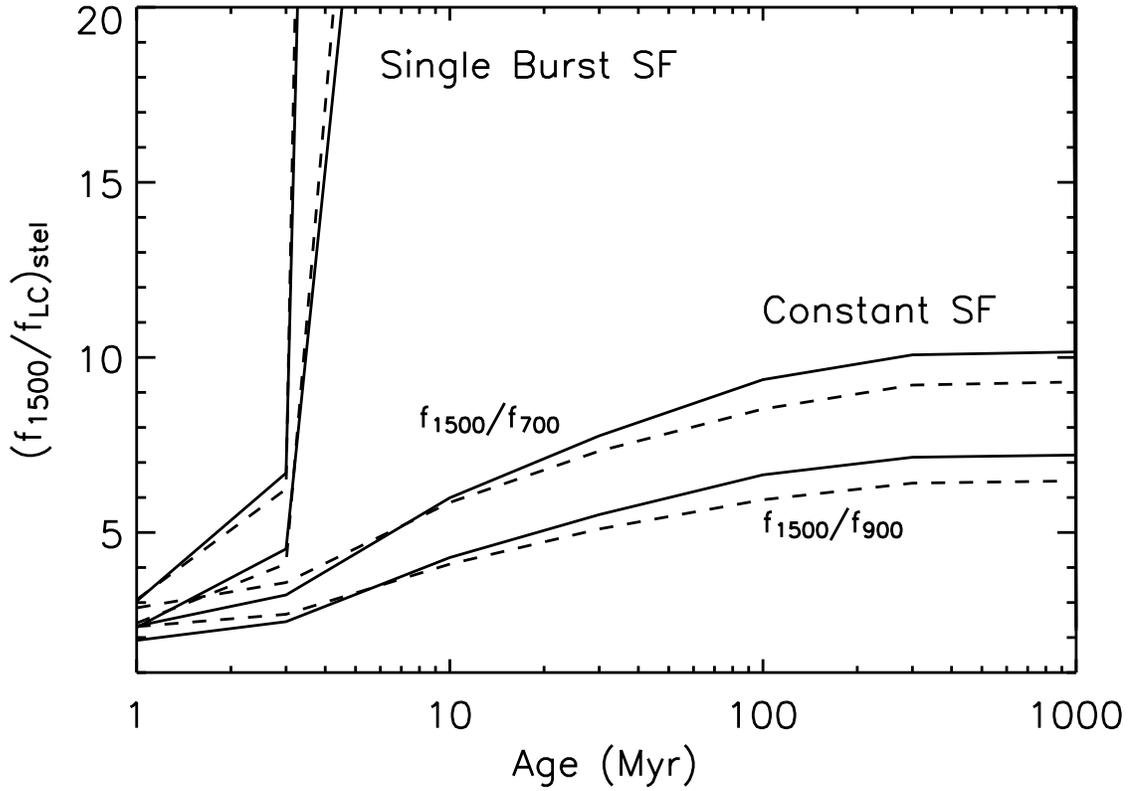}
\caption{The $f_{1500}/f_{700}$ and $f_{1500}/f_{900}$ intrinsic flux ratios as a function of time since onset of star-formation using the BC03 (solid) and Starburst99 (dashed) models . \label{fig:age_ratio}
}
\end{figure}

Since the lifetimes of the O-stars (which emit the Lyman continuum flux) are so much shorter than the B- and A-stars that dominate the 1500\AA\ flux emission, the $f_{1500}/f_{700}$ flux ratio is highly dependent upon the starburst age and the star-formation history.  In Figure \ref{fig:age_ratio} we plot the expected flux ratio as a function of age for both the BC03 and Starburst99 models with single burst and constant star-formation histories.  In the single burst systems, the dying O-stars are not replenished with new star-formation, causing the $f_{1500}/f_{700}$ ratio to increase rapidly after a few Myrs.  In the constant star-formation scenario, early in the burst, the dying O-stars are being replenished, while the B- and A-stars accumulate, slowly increasing the $f_{1500}/f_{700}$ ratio.  Eventually, the B- and A-stars begin to die as well and no longer increase in number.  Thus, the flux ratio remains constant after $\sim$300 Myr.  The flux ratio of a starburst with declining star-formation rates evolves between these two extremes.  If the observed galaxy is undergoing a secondary burst, there may be significant remaining flux at 1500 \AA\ but only if the bursts are separated by less than a few 100 Myr.  \citet{shapley01} find that constant star-formation scenarios fit well to the Lyman Break galaxy population with a median age, $t_{sf}=320$Myr.  We will show in Section \ref{comp_sed} that we obtain similar star formation histories for our sample.  Therefore, we expect the typical intrinsic break amplitude to be $(f_{1500}/f_{900})_{stel} \sim 6$ or $(f_{1500}/f_{700})_{stel} \sim 8$.

\subsubsection{Initial Mass Function}
The slope of the high-mass end of the stellar initial mass funtion (IMF) determines the relative number of O-stars to B- and A-stars, and therefore impacts the $f_{1500}/f_{700}$ ratio.  The Salpeter IMF \citep{salpeter55}, commonly used in studies of high-redshift galaxy formation, has a mass distribution characterized by $\alpha = -2.35$, where $\xi(M) \propto M^{\alpha}$.  Many studies of the high-mass end of the IMF agree with this IMF slope, though with significant scatter between star clusters \citep{massey98,scalo98,kroupa01}.  No direct measurements of the IMF can be done at high redshift, but initial indications from model fits to the UV spectra of a lensed LBG (MS1512-cB58) indicate that the high-mass IMF is nearly Salpeter and must extend above 50 M$_{\odot}$ \citep{pettini00}.  Given the agreement of detailed local studies and initial indications at high-redshift, we use a Salpeter IMF in our analysis.  It should be noted, however, that several studies suggest that top-heavy IMFs in some high-redshift star-formation may offer a possible explanation of $\alpha$-element abundances seen in present-day cluster ellipticals \citep{faber92,worthey92,matteucci94,gibson97,thomas99,nagashima05}.  A top-heavy IMF would significantly decrease the expected, UV-to-LC ratio expected with a Salpeter IMF.

\subsubsection{Metallicity}

The metallicities of galaxies such as those in our sample as well as LBGs, range from $Z \sim 0.2-1.0 Z_{\odot}$.  Over this range, the amplitude of the Lyman break does not change significantly \citep[see eg. Fig. 75,][]{leitherer99}.  In our subsequent analysis we use solar metallicity models. 

\subsection{IGM absorption}
\label{igm}
\begin{figure}
\epsscale{1.0}
\plotone{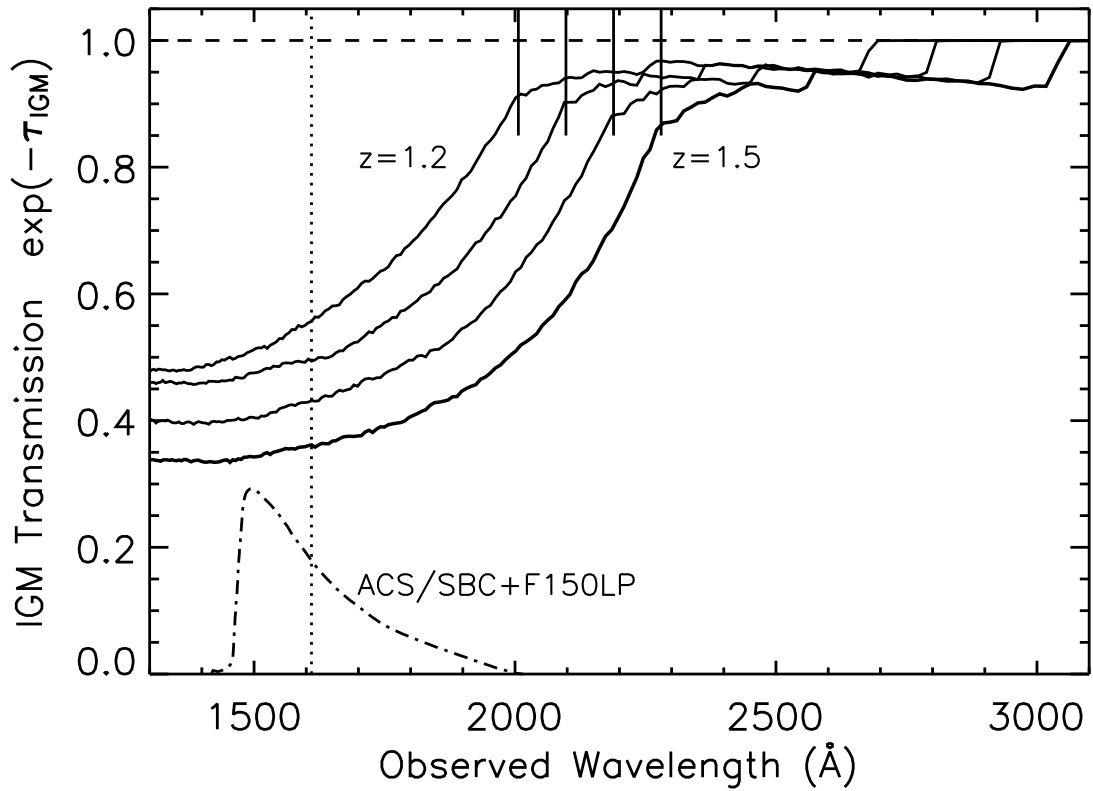}
\caption{The average transmission through the IGM of UV light from galaxies at redshift 1.2, 1.3, 1.4, and 1.5 (left to right).  Each curve is computed from simulations along 1000 lines of sight through the Ly$\alpha$ forest.  The corresponding Lyman limits are denoted by solid vertical lines.  As a reference, the transmission curve is plotted for the ACS/SBC F150LP filter used in our observations (dot-dashed line).  The vertical dotted line is the pivot wavelength of the filter curve. \label{fig:avg_igm}}
\end{figure}

\begin{figure}
\epsscale{1.0}
\plotone{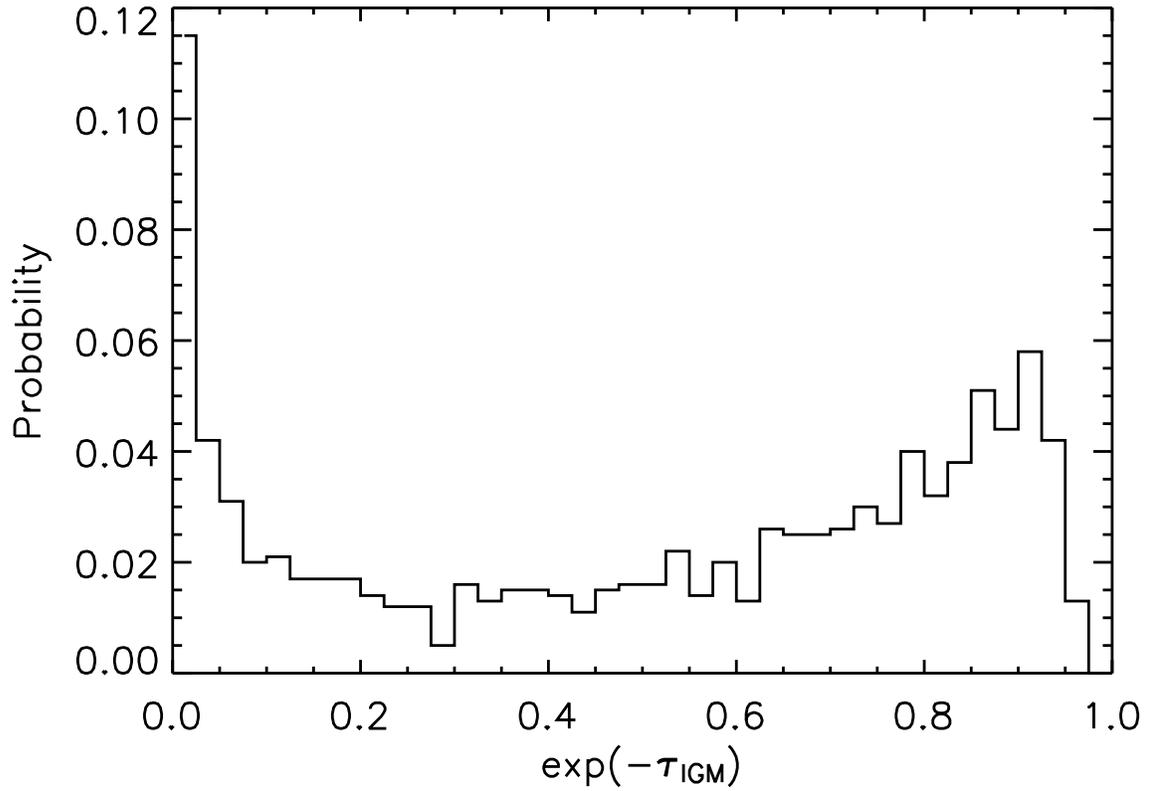}
\caption{The distribution of IGM transmission within the F150LP filter for 1000 lines of sight towards galaxies at $z=1.3$.  The spike at zero transmission is due to Lyman Limit Systems (LLSs) and Damped Ly$\alpha$ systems (DLAs) within $\Delta z \sim 0.4$ from the target galaxy.  The mean and median are 0.51 and 0.59, respectively. \label{fig:igm_hist}}
\end{figure}

Along the line of sight (LOS) to any high-redshift galaxy, there are hundreds of intervening neutral hydrogen clouds whose Lyman line and continuum absorption significantly affects the measured far-UV fluxes of the galaxies.  The column density and redshift distributions of these HI absorbers have been well studied and have been used to model the transmission through the IGM as a function of the observed galaxy's redshift \citep{madau95, bershady99}.  We make use of the best available data on the column density and redshift distributions of the Ly$\alpha$ Forest \citep{kim97}, Lyman Limit Systems \citep[LLSs,][]{storrie-lombardi94} and Damped Ly$\alpha$ systems \citep[DLAs,][]{storrie-lombardi00} to simulate the effects of IGM Lyman line and continuum absorption on the observed fluxes of galaxies at $z\sim1$.  We simulate 1000 sight lines to galaxies at redshift intervals $\Delta z = 0.05$ between $1.0<z<1.5$.  For each sight line we randomly place absorbers with column densities and redshifts which are consistent with their empirical distributions.  We note that this does not account for clustering of the absorbers, which may be important, especially near the observed galaxy if it is within an overdensity.  We then compute the absorption line profiles (velocity width $\beta = 30$ km s$^{-1}$) and apply continuum absorption for each absorber to compute the total transmission through the IGM as a function of wavelength (See Siana et al. 2007 for more details).  The mean transmission from 1000 LOSs is plotted in Figure \ref{fig:avg_igm}.  At the redshifts of interest, the average transmission of UV light changes from 0.57 to 0.37 between $1.2<z<1.5$.  Therefore, our measurements are more sensitive at lower redshifts as less of the Lyman continuum is being absorbed.  However the distribution function of the IGM transmission is highly non-Gaussian.  Figure \ref{fig:igm_hist} shows the distribution of IGM transmission (as measured through our filter), indicating an almost bimodal distribution.  About 20\% of the simulated lines-of-sight contain large column density absorbers, resulting in very little transmission ($<5$\%).  Conversely, about 40\% of the LOSs have no intervening large column density absorbers and therefore have greater than 60\% transmission.  Though the mean and median are similar (0.51 and 0.59, respectively at $z = 1.3$), it is clear that the absorption correction will change significantly between galaxies.  This distribution makes it difficult to draw conclusions from small samples if one can't determine the properties of the intervening IGM.  We will use this IGM transmission distribution in a Monte-Carlo simulation to interpret our measurements (see Section \ref{z1_comp}). 

\subsection{Dust}
\label{dust}
\begin{figure}
\epsscale{1.0}
\plotone{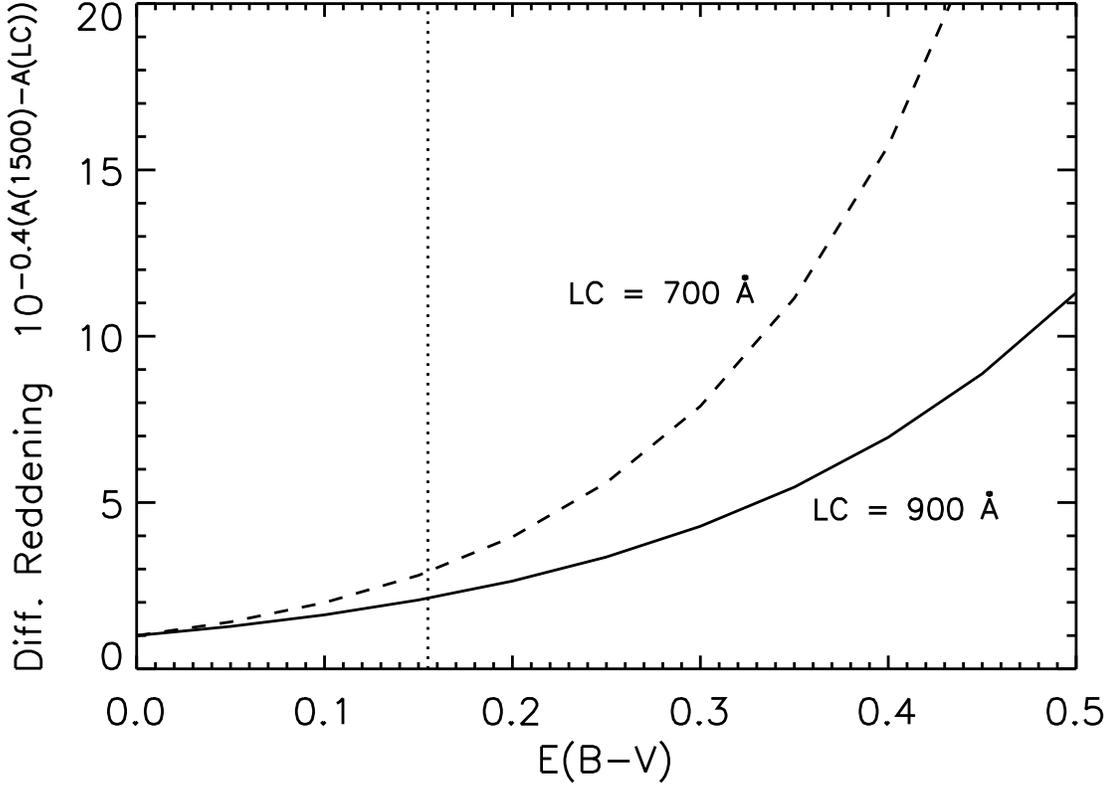}
\caption{The differential reddening from 1500 \AA\ and the Lyman continuum (defined as $10^{-0.4(A(1500)-A(LC))}$ where LC is the either 900 or 700 \AA) as a function of dust reddening for a Calzetti reddening law extrapolated to $\lambda=700$ \AA.  The dotted vertical line shows the median value of $E(B-V)=0.155$ for \citet{shapley01} for Lyman Break Galaxies derived from broadband photometry.  The tail of the reddening distrubution for LBGs extends out to $E(B-V)=0.4$.  The 700 \AA\ flux is affected far more than at 900 \AA\ showing that our measurements are more sensitive to dust than those measurements just blueward of the Lyman limit.  \label{fig:dust_ratio}
}
\end{figure}

Although it is not necessary to know the level of dust attenuation to determine $f_{esc,rel}$, it is important for converting between the relative and absolute escape fractions and in determining whether HI or dust is primarily responsible for the Lyman continuum absorption.  If the dust extinction continues to increase at $\lambda < 912$\AA\, then the measured ratio of UV-to-LC flux will be even larger than the intrinsic Lyman Break of the stellar SEDs.  The level of increase is defined by the amount of dust attenuation and the reddening law.  Of course, the extinction curve below the Lyman limit is impossible to measure due to HI absorption, but theoretical models of the ISM dust composition and size distribution predict extinction curves that continue to rise to $\lambda \sim 700$ \AA\ \citep[see Figure 14 in][]{weingartner01}.  In Figure \ref{fig:dust_ratio} we plot the differential reddening across the Lyman Break as a function of optical color excess, E(B-V), for a Calzetti reddening law \citep{calzetti97}.  The reddening law at $\lambda<1200$ \AA\ is extrapolated with a constant slope ($dk(\lambda)/d\lambda$, where $k(\lambda) = E(B-V)/A_{\lambda}$) derived at $1100<\lambda<1200$ \AA, approximating the increasing extinction predicted below the Lyman limit in \citet{weingartner01}.  We note that while this extrapoloation is a reasonable assumption at 900\AA\ (extinction curves in SMC, LMC, MW all continue to rise at shorter wavelengths up to the Lyman limit), the extrapolation is somewhat uncertain at $700$ \AA.  The level of reddening in LBGs typically varies from $0.0<E(B-V)<0.4$ with a median value of $E(B-V)\sim0.15$ \citep{papovich01,shapley01}.  Therefore, the dust alone can cause the UV-to-LC ratio to vary widely over this range (additional factors of 1-8 at 900 \AA\ and 1-20 at 700 \AA) with typical values of the differential dust attenuation of $\sim2$ and $\sim3$ at 900 and 700 \AA, respectively.  At color excesses of $E(B-V) > 0.5$, we can not detect the Lyman continuum at 700 \AA\ in this study.

\section{Observations}
\label{observations}
\begin{figure}
\epsscale{1.0}
\plotone{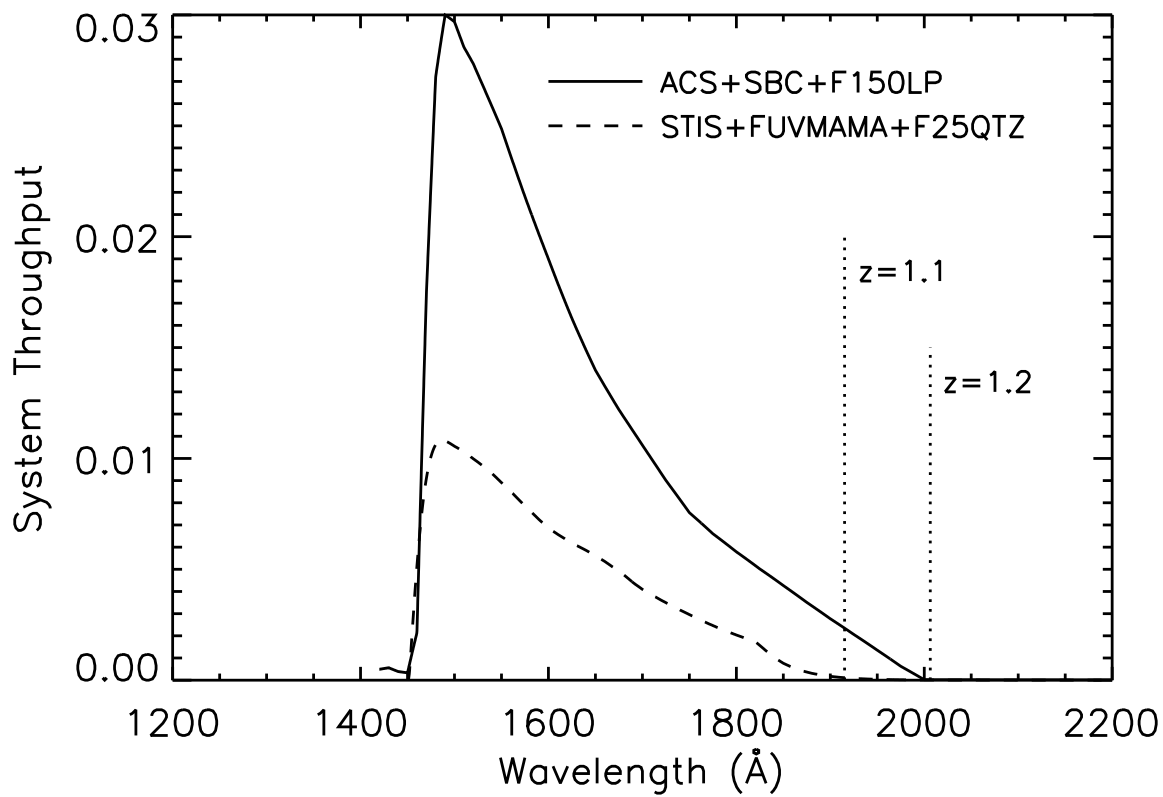}
\caption{Total system throughput for the ACS+SBC+F150LP (solid line) and STIS+FUV-MAMA+F25QTZ (dashed) configurations.  The ACS/SBC is nearly 3x more sensitive than the STIS configuration.  The dashed vertical lines denote the location of the Lyman Limit at z=1.1 and z=1.2, the low redshift cutoffs for our sample selection for each configuration. \label{fig:filter_curves}}
\end{figure}

We have obtained far-UV imaging of the Hubble Deep Field North and the Hubble Ultra Deep Field at 1600 \AA\ in three HST General Observer programs (7410, 9478, 10403).  The HDF-N was observed with STIS through the FUVQTZ filter covering 1.02 arcmin$^2$ \citep{gardner00b}.  \citet{teplitz06} imaged most of the rest of the HDF-N (3.77 arcmin$^2.$) with the F150LP filter on the Solar Blind Channel of the ACS.  These two instrument configurations have very similar throughputs (see Figure \ref{fig:filter_curves}), with a lower wavelength cutoff at $\lambda \sim 1450$ \AA\ and a decreasing sensitivity out to 2000 \AA\ (1840 \AA\ for STIS).  The effective wavelengths for the STIS and SBC filters are $\lambda = 1600$\AA\ and $\lambda = 1610$ \AA, respectively.  The HUDF observations (Siana et al. 2007, in prep) use the same ACS/SBC configuration as in the HDF-N and cover 7.77 arcmin$^2$, nearly the same area as the NICMOS HUDF treasury program \citep{thompson05}.

\begin{figure}
\epsscale{1.0}
\plotone{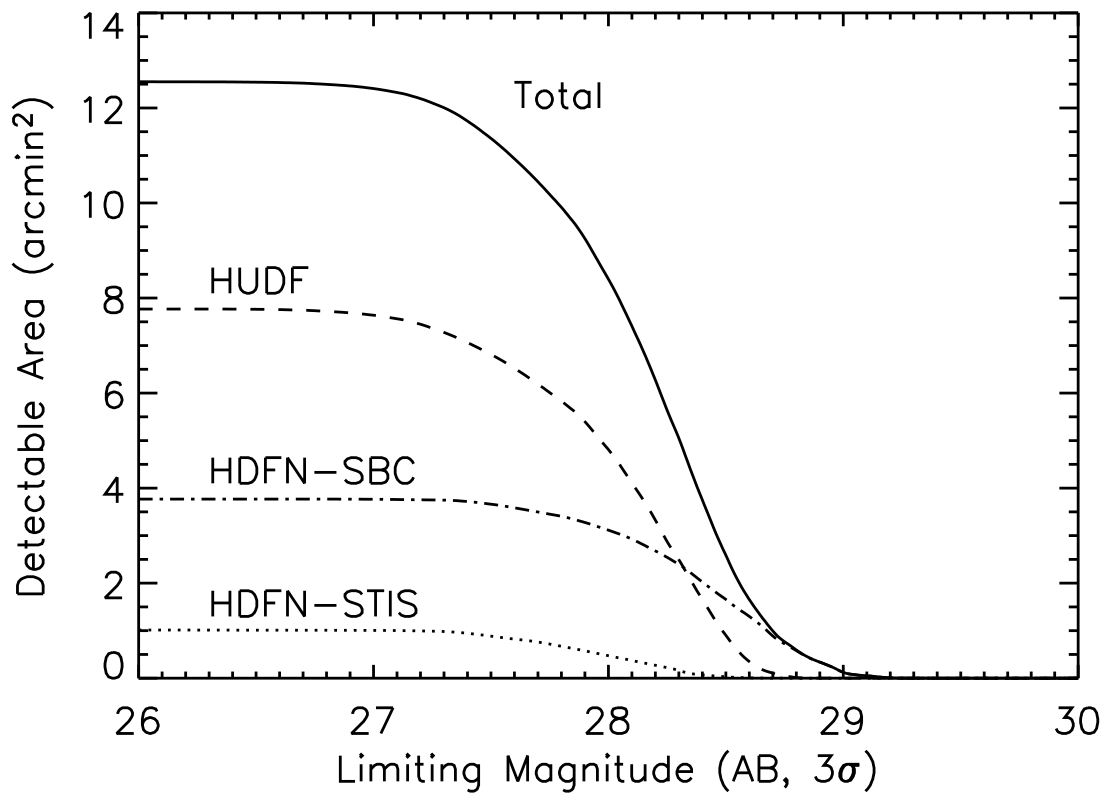}
\caption{Detectable area as a function of limiting magnitude (AB, 3$\sigma$) within a 1$''$ diameter circular aperture.  The solid line is the total area from all three surveys.  The dashed, dotted, and dash-dotted lines are the areas of the HUDF, HDFN-STIS, and HDFN-SBC surveys respectively.  The HUDF survey is much larger, allowing us to detect more objects, while the HDF-SBC survey is deeper.  We are are sensitive to fainter galaxies if their extraction isophotes are smaller.  \label{fig:sens_hist}}
\end{figure}

Both the SBC and STIS use Multi-Anode Microchannel Arrays (MAMAs), which have no read noise and are insensitive to cosmic rays.  The primary source of noise is dark current, which has two components.  The first component is a fairly uniform count rate that doesn't change with the detectors' temperature.  The second component is a temperature-dependent ``glow'' which arises at T $>25$ C and is near the center of the SBC MAMA and one corner of the STIS MAMA.  As the instrument warms up, the dark current from the ``glow'' increases, decreasing the sensitivity of the observations.  Because of this, the images taken at the end of a series of observations are significantly less sensitive than those at the beginning.  Therefore, the dark current changes substantially from pointing to pointing, resulting in significantly different sensitivities across the field.  In addition, there is significant area where adjacent frames overlap.  Weight maps have been defined which account for the total exposure time and dark count per pixel so that accurate sensitivities can be computed.  The dark subtraction and weight map production used the same procedures as in \citet{teplitz06}.  Figure \ref{fig:sens_hist} shows the cumulative area for which we expect to obtain a $3\sigma$ detection of a galaxy of a given magnitude within a 1$''$ diameter aperture.   

\subsection{Galaxy selection}
In order to avoid contamination from photons with $\lambda_{rest} > 912$\AA, we choose galaxies at redshifts large enough to shift the Lyman Limit to wavelengths at which our filter/detector configurations are no longer sensitive.  This occurs at $\lambda_{obs} > 1915$\AA\ ($z>1.1$) for the STIS observations and $\lambda_{obs} > 2000$\AA\ ($z>1.2$) for the SBC observations.  The MAMA detectors are sensitive to wavelengths longer than $\lambda > 2000$\AA\ but their quantum efficiency is $10^{-3}$ times lower at 2000 \AA\ than at 1500\AA\ and decreases rapidly toward redder wavelengths.

Our sensitivity to Lyman continuum radiation falls off rapidly for galaxies with $z>1.5$ because 1) only the youngest, least reddened galaxies will have significant flux at $\lambda<600$\AA, 2) the rest-frame Lyman continuum of the galaxy will be diminished by continuum absorption from an increasing number of neutral HI clouds along the line-of-sight to the galaxy.  We therefore only choose galaxies with $z \leq 1.5$.  The spectroscopic redshifts are primarily from the Team Keck Redshift Survey \citep[TKRS][]{wirth04} in the HDF-N and the VLT/FORS2 spectroscopy \citep{vanzella05,vanzella06} in the HUDF, with a few additional redshifts from \citet{cowie04} and \citet{lefevre04}.  In addition to these redshift requirements, we also ensure that there are no sources with optical spectra with high-ionization emission lines (NeV,NeIII) or with strong nuclear point sources indicative of strong AGN activity.  Finally, we removed the three reddest ($B-V>0.6$) galaxies from our sample as their SEDs are indicative of older stellar populations with little ongoing star-formation.  The 8 galaxies in the HDF-N and 13 galaxies in the HUDF that meet these requirements are listed in Table \ref{tab:results}.

Recent measurements have shown that the total optical throughput of the SBC at red wavelengths is much higher than was measured before launch ($3.4 \times 10^{-6}$ vs. $9.5\times10^{-9}$ at $\lambda = 3500$\AA, STScI Analysis Newsletter, Nov. 30, 2006\footnote{http://www.stsci.edu/hst/acs/documents/newsletters}).  The total system throughput at 3500 \AA\ is still four orders of magnitude less than at 1500 \AA, but we must be careful that the `leaking' optical light does not significantly affect our sensitive far-UV measurements.  As a test, we examined the effects of the red leak with two types of galaxies: a `typical' galaxy ($t_{sf}=300$ Myr, $E(B-V)=0.2$) and a `worst-case' (ie. extremely red) galaxy ($t_{sf}=1$ Gyr, $E(B-V)=0.5$).  We shifted the galaxy templates to $z=1.2$, applied average IGM absorption, and determined the ratio of leaking flux density (with $\lambda_{rest}>912$ \AA) to the measured flux density at $\lambda_{rest} > 1500$ \AA.  These ratios were 0.003 and 0.002 for the worst-case and typical galaxies, respectively.  Therefore, we can be confident that our results will not be significantly affected if our observed flux ratios are significantly above $f_{700}/f_{1500} \geqsim 0.003$.  We show in Section \ref{ratios} that our most sensitive limits are $f_{700}/f_{1500} \geqsim 0.01$, and should therefore be unaffected by this leaking optical light.

\subsection{Photometry}
For the optical and near-IR photometry, we used the NICMOS-selected photometric catalogs of \citet{dickinson00} in the HDF-N and \citet{thompson05} in the HUDF.  The HUDF near-IR photometry was corrected by $\Delta J=0.30$ and $\Delta H = 0.18$ as discussed in \citet{coe06}.  For both of these catalogs, source detection and isophotal apertures were determined from the summed J (F110W) and H (F160W) band NICMOS images.  These same apertures were used to extract fluxes from the optical images as well.  Each isophote was inspected by eye to ensure there was no contamination from nearby galaxies and that no galaxy was improperly fragmented into several pieces.  

The UV photometry was extracted using isophotes defined by the optical B-band ($\lambda_{rest} \sim 1900$\AA), which should directly trace new star-formation at these redshifts.  Here we make the reasonable assumption that significant far-UV signal will not be detected where there is no measurable flux in the deeper B-band images.  Because the B-band images are so sensitive, their faintest isophotes have large areas which contain very little flux.  We find that decreasing the aperture to exclude the faintest 20\% of the galaxy reduces the isophotal areas by factors of $\sim$2-3.  Therefore,  in order to maximize the signal-to-noise of the photometry, we use as our extraction aperture the isophote which contains only 80\% of the B-band flux, increasing our sensitivity by $\sim$ 0.4-0.6 mags.  No correction for foreground extinction is necessary because both the HDF-N and HUDF have clear lines of sight through the galaxy \citep[$E(B-V) \sim 0.01$,][]{schlegel98}.  

As discussed by \citep{gardner00b,brown00,teplitz06}, the dark current is the principle source of noice in these observations.  The individual frames were weighted by the square of the exposure time and divided by the total dark (primary + ``glow'').  As the dark count scales with exposure time, these weight maps scale linearly with the ratio of exposure time to dark rate.  Thus, the final weight maps are the square of the signal-to-noise ratio for objects fainter than the background.  For non-detections, we derive 3$\sigma$ upper limits from these weight maps. 

\section{Results}

\begin{figure}
\epsscale{1.0}
\plotone{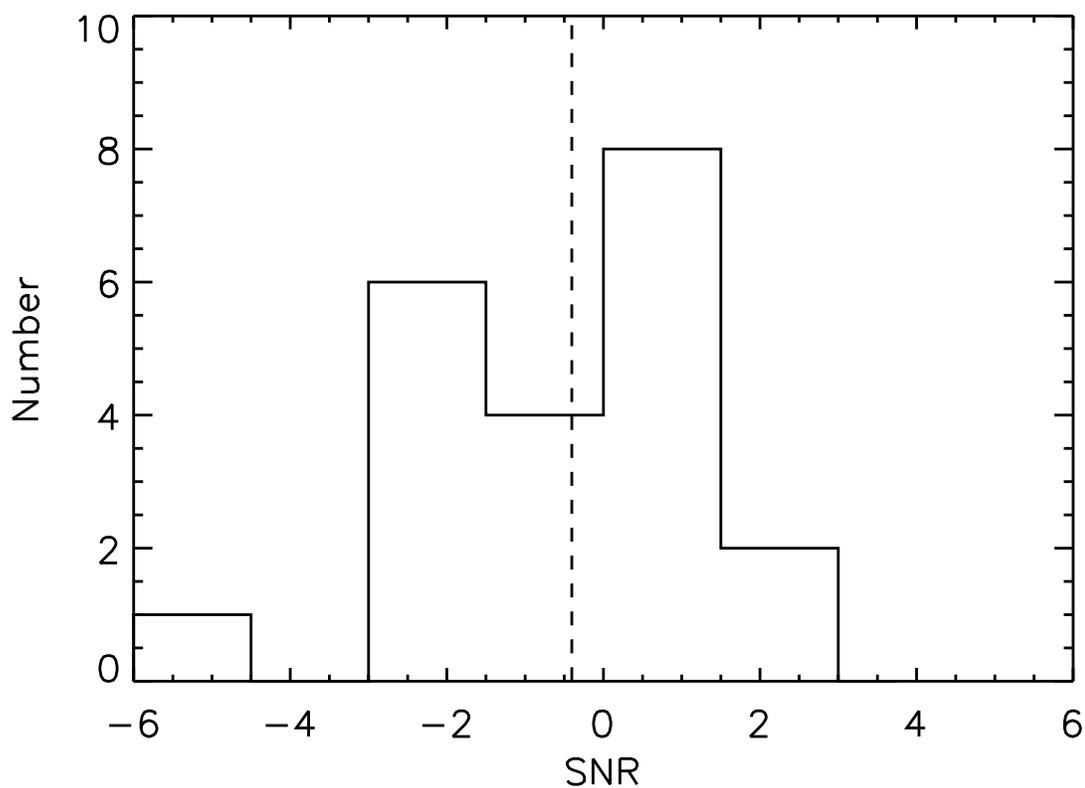}
\caption{Histogram of measured signal-to-noise ratios for the 29 galaxies in our sample.  There are no detections above $SNR>3$.  The dashed vertical line is the average $<SNR> = -0.40$.  The standard deviation of this distribution, $\sigma = 1.7$, is slightly larger than expected.  This is due to small errors in ``background'' subtraction of a few sources (namely the object at $SNR=-5$) since the dark current is non-planar. \label{fig:snr_hist}} 
\end{figure}

We do not detect any of the 21 galaxies in our sample (with $S/N > 3$) and the $3\sigma$ upper limits to the far-UV flux are given in Table \ref{tab:results}.  The distribution of measured $SNR$ (See Figure \ref{fig:snr_hist}) is centered around zero as expected.  There is one galaxy which has a large negative flux, but that is due to poor background subtraction due to edge effects.  The other sources are not affected by this.  For the bluer, more luminous galaxies in our sample, we expect to derive significant limits to the escape fraction of ionizing photons.  In Section \ref{ratios} we derive $f_{esc,rel}$ limits by reproducing the techniques of previous high-redshift escape fraction studies.  Specifically, we calculate flux ratios on either side of the Lyman limit and compare to the average intrinsic values to deduce a relative escape fraction.  In Section \ref{sed} we fit SED models to ascertain dust and age parameters for each galaxy to better determine the intrinsic Lyman continuum flux and thus the relative escape fraction.

\subsection{Flux Ratios}
\label{ratios}

To replicate the same methods used in previous high-redshift escape fraction studies, the 3$\sigma$ limits to the $(f_{1500}/f_{700})_{obs}$ flux ratio were computed and are listed in Table \ref{tab:results}.  The $f_{1500}$ value is computed by interpolating the optical photometry and the $f_{700}$ limit is derived from the weight maps which account for total exposure time and dark current per pixel.  The flux ratios are converted to $f_{esc,rel}$ with Eqn \ref{eqn: fesc_rel}.  We choose to use $(f_{1500}/f_{700})_{stel} = 8$ for our intrinsic stellar flux ratio since the ratio varies from 6-10 between $0.01<t_{age}<0.3$ Gyr.  We use the redshift dependent average IGM absorption correction $exp(-\tau _{IGM}(z))$ from our simulations defined in Section \ref{igm}. 

\begin{figure}
\epsscale{1.0}
\plotone{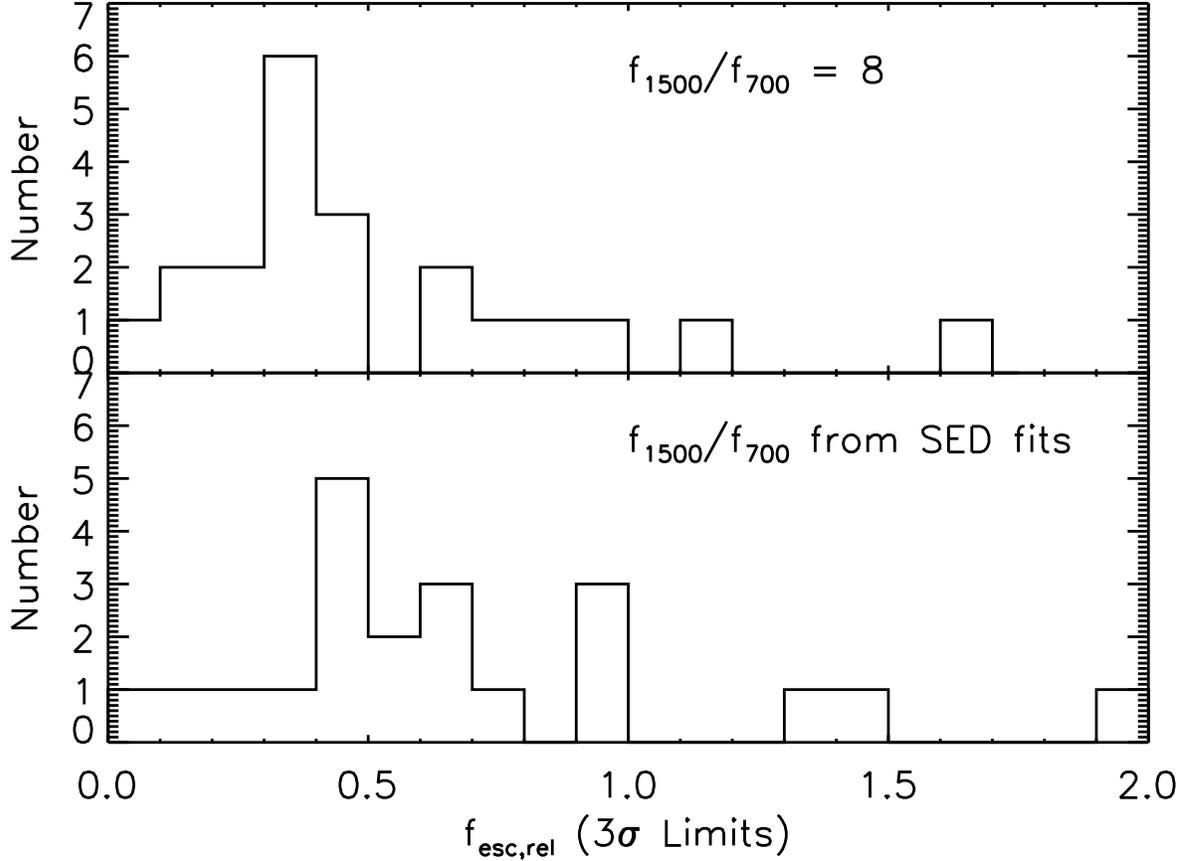}
\caption{Histogram of the $f_{esc,rel}$ limits of our sample. The top panel limits are derived assuming a UV-to-LC ratio $(f_{1500}/f_{700})_{stel}=8$, whereas the lower panel show the limits derived when taking the UV-to-LC ratio from the best fit model.  Limits of $f_{esc,rel} < 1.0$ indicate additional attenuation (by either dust or HI) in addition to the attenuation at 1500 \AA.  Eighteen (of 21) galaxies have limits of $f_{esc,rel} \leq 1.0$ and 9 have $f_{esc,rel} < 0.5$ (when using SED fits).  \label{fig:fesc_hist}}
\end{figure}

The distribution of $f_{esc,rel}$ limits are plotted in Figure \ref{fig:fesc_hist} (top panel).  Nineteen of the galaxies have $f_{esc,rel} \leq 1.0$, indicating either increased dust attenuation or HI absorption of the Lyman continuum.  

\subsection{SED fitting}
\label{sed}
In addition to deriving flux ratios, we have fit SED models to the high S/N optical/near-IR photometry.  In this way, we derive better estimates of the starburst age and dust reddening for each galaxy, rather than assuming average values for the entire sample.  We used all available optical/near-IR photometry from the deep HST images for the SED fits.  These data, in addition to measuring the slope of the UV continuum, span the 4000 \AA\ break, allowing us to break the degeneracy of age and dust effects on the UV slope.  The observed far-UV (1600 \AA) limits are not used in the fits of the SEDs.  

We fit a suite of nine Bruzual-Charlot models with different starburst ages varying logarithmically between $0.001<t_{age}<10$ Gyr and allow dust attenuation \citep[Calzetti Law,][]{calzetti97} to vary as a free parameter.  The metallicity was fixed at the solar value and the IMF is assumed to be Salpeter \citep{salpeter55}.  We attempt to fit models with instantaneous starbursts, exponentially declining star formation with an e-folding time $\tau = 100$ Myr and constant star-formation models. 

\begin{figure}
\epsscale{1.0}
\plottwo{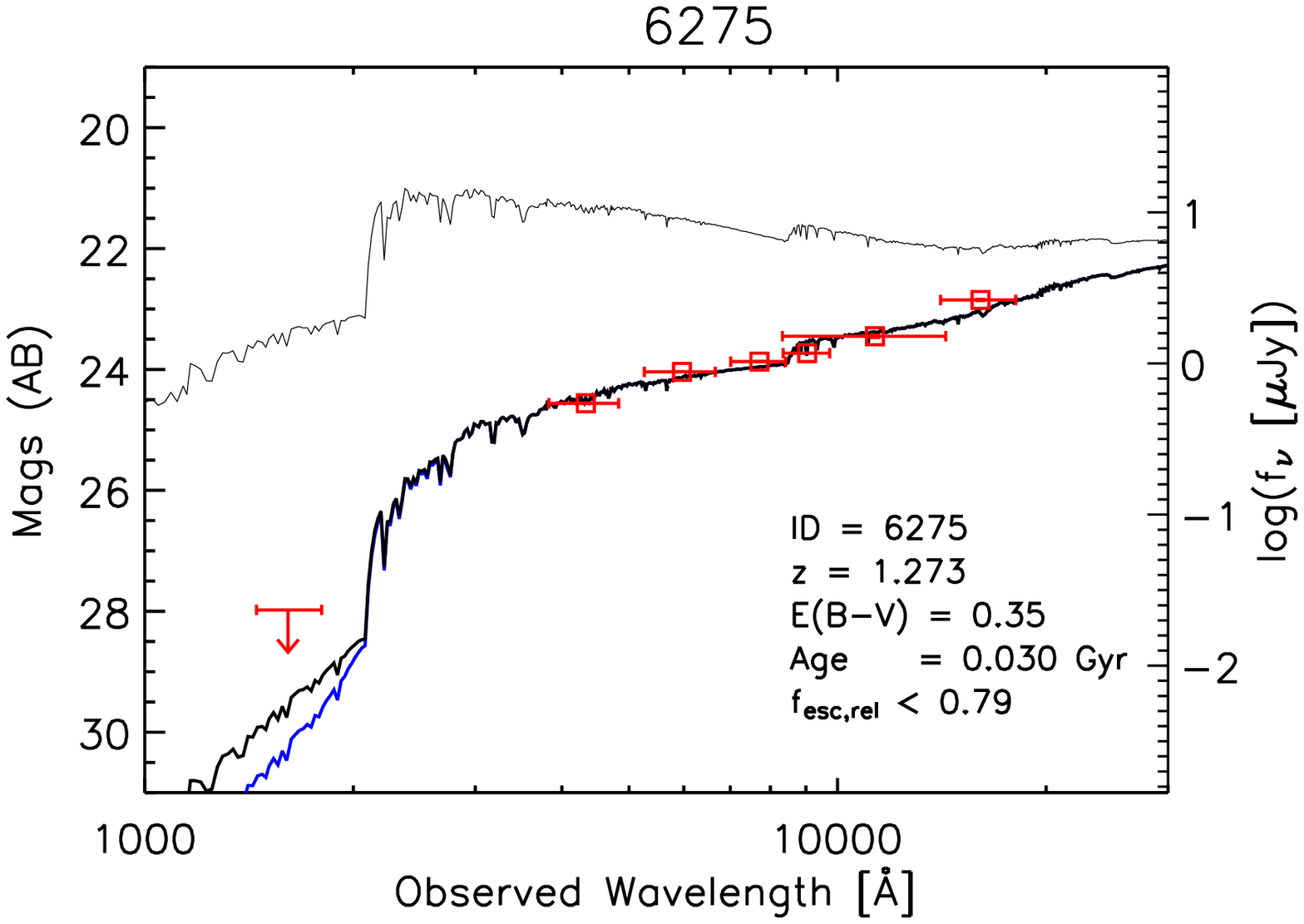}{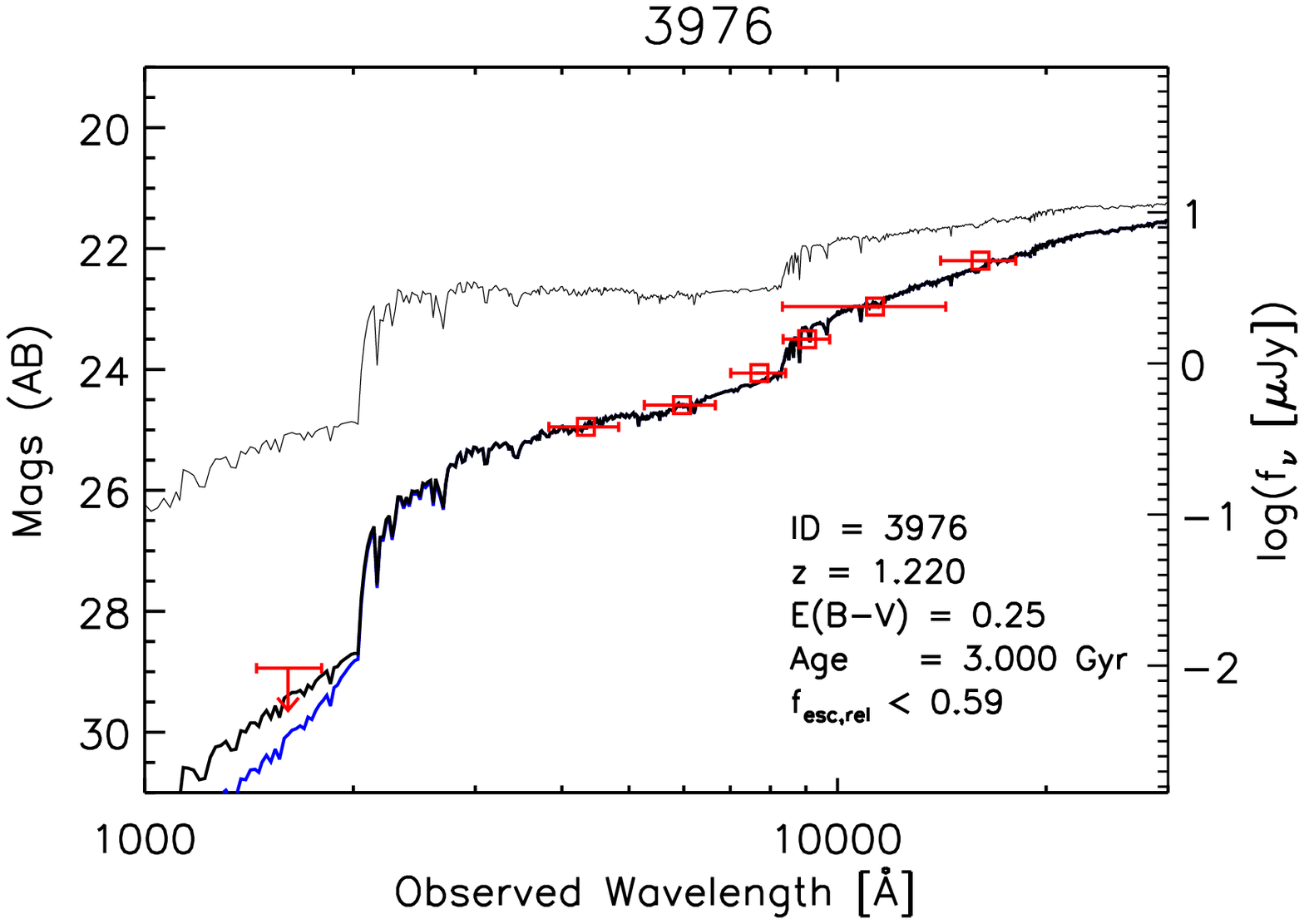}
\epsscale{0.45}
\plotone{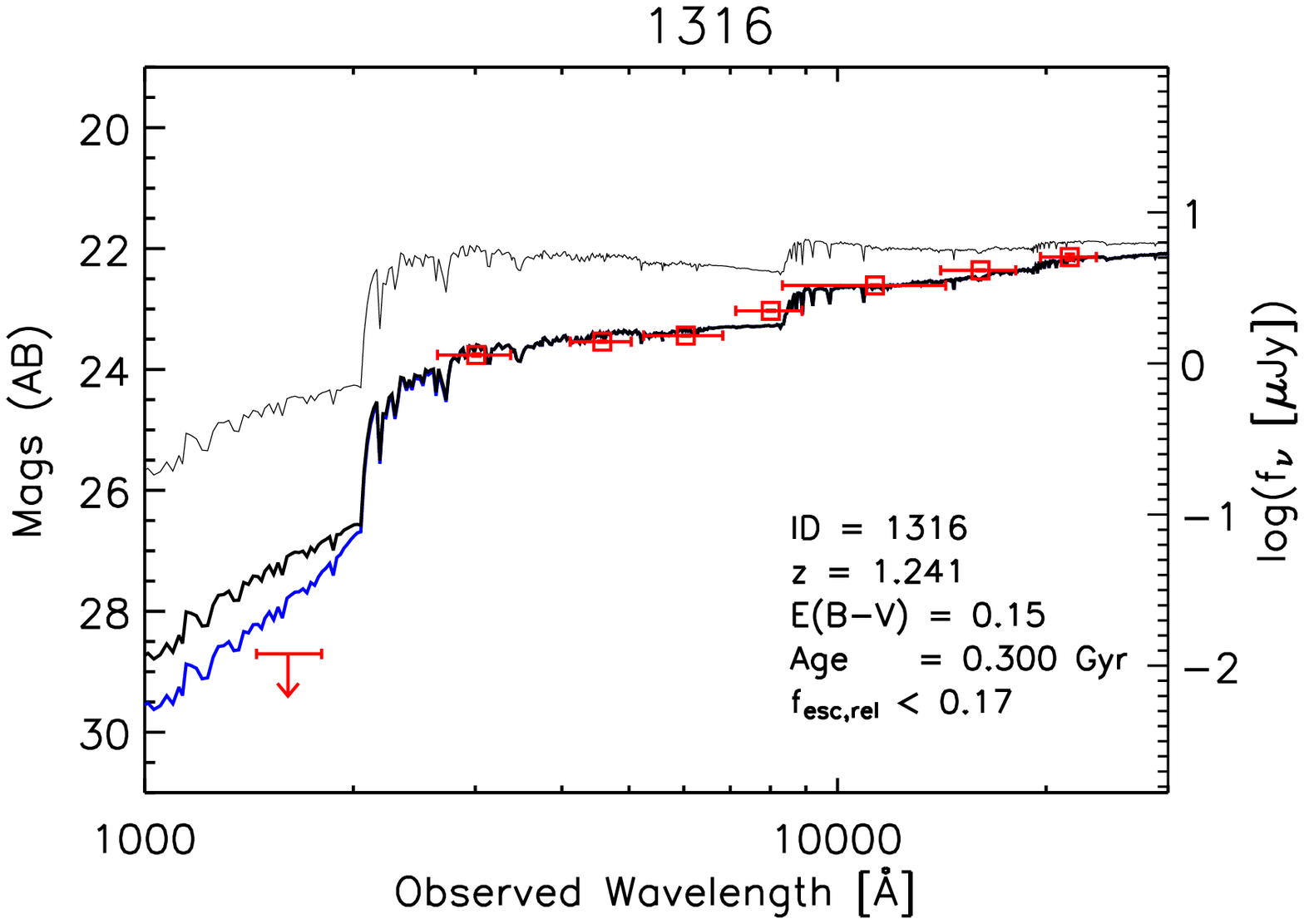}
\caption{Best fit SEDs for three galaxies.  Plotted in each panel are the original SED (thin), with best-fit reddening applied (thick), and with average IGM absorption applied (blue).  Top Left:  This galaxy (J033237.07-274617.3) has no indication of a Balmer break but has a red spectral slope so it is fit with a young (30 Myr) and dusty (E(B-V)=0.35) model  Top Right:  A red galaxy (J033244.16-274729.5) with a prominent Balmer break fit with a very old (3 Gyr), relatively dusty (E(B-V)=0.25) template.  Bottom:  A galaxy (J123643.41+621151.6) with an SED very similar to typical LBGs (Age = 300 Myr, E(B-V)=0.15). The far-UV 3$\sigma$ limits are plotted with downward arrows.  Note that our ability to detect leaking Lyman continuum is largely dependent upon the level of dust attenuation. \label{fig:sed_examples}}
\end{figure}

The results of the fits are given in Table \ref{tab:results} and a few examples are plotted in Figure \ref{fig:sed_examples}.  We find that the best fit model is a constant star-formation model for 90\% (19 of 21) of the galaxies, confirming that these galaxies, which are bright and blue in the rest-frame ultraviolet, are actively undergoing star-formation.  The fits to the two other galaxies have $\chi^2$ values that are not significantly better than those of the constant star-formation model.  Therefore, we have chosen to list only the contant star-formation fits in Table \ref{tab:results} and use only these models in the subsequent analysis.

The distribution of relative escape fractions using the ($f_{1500}/f_{700})_{stel}$ from the best-fit model is shown in Figure \ref{fig:fesc_hist} (bottom panel).  Eighteen of the galaxies have a relative escape fraction less than unity.  We stacked the fluxes at 1500 \AA\ and added the 700 \AA\ errors in quadrature to achieve a summed $3\sigma$ limit $f_{esc,rel} < 0.08$ at 700 \AA, though most of the signal comes from our brightest source.  

As a consistency check, we also fit to the Starburst99 \citep{leitherer99} models with constant star-formation, solar metallicity, and Salpeter IMF.  There were small differences in age and dust determinations, but no systematic difference from the results derived with the Bruzual \& Charlot models.

\subsection{Individual Galaxies}

\subsubsection{J123652.69+621355.3}
\begin{figure}
\epsscale{1.0}
\plottwo{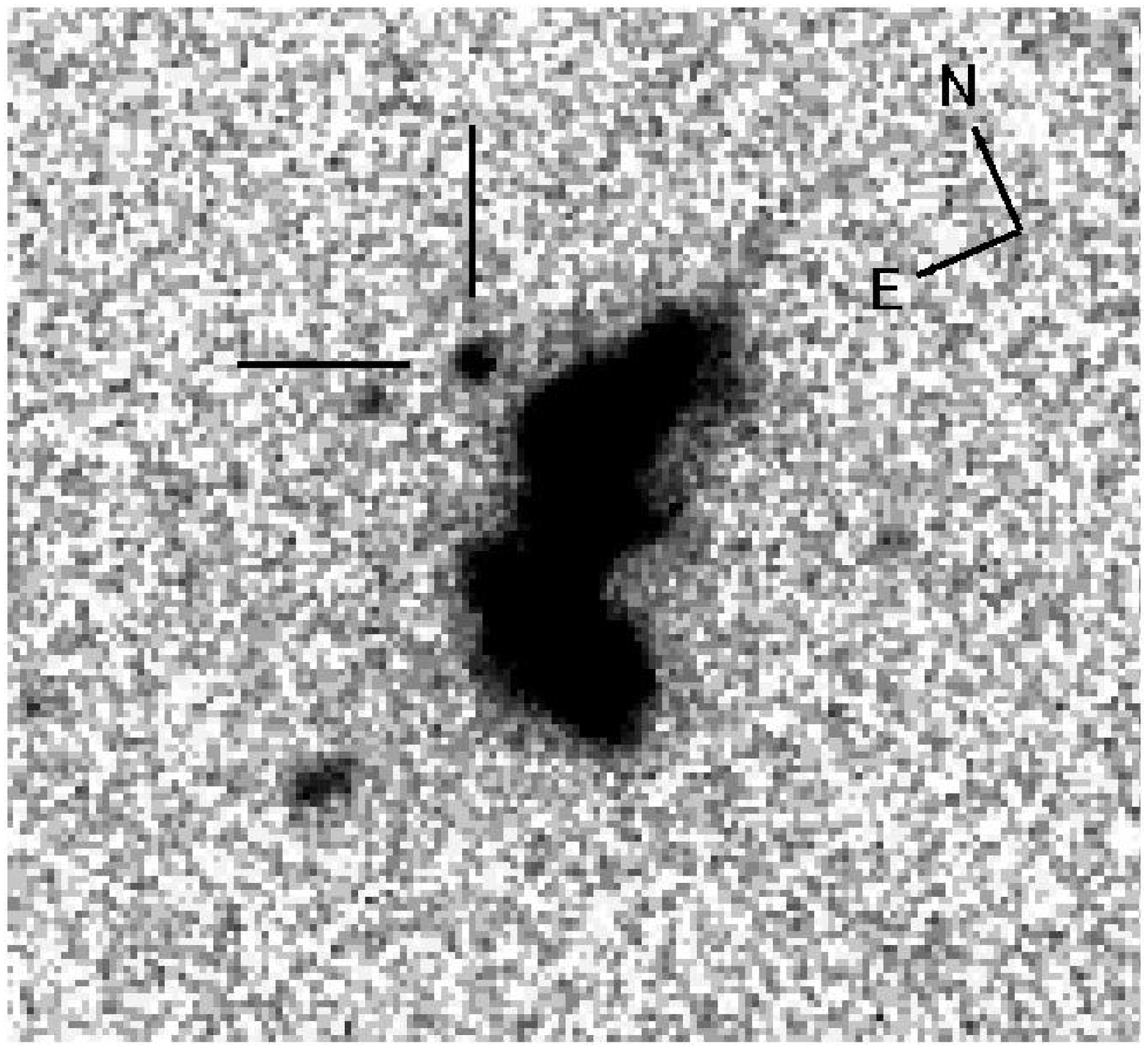}{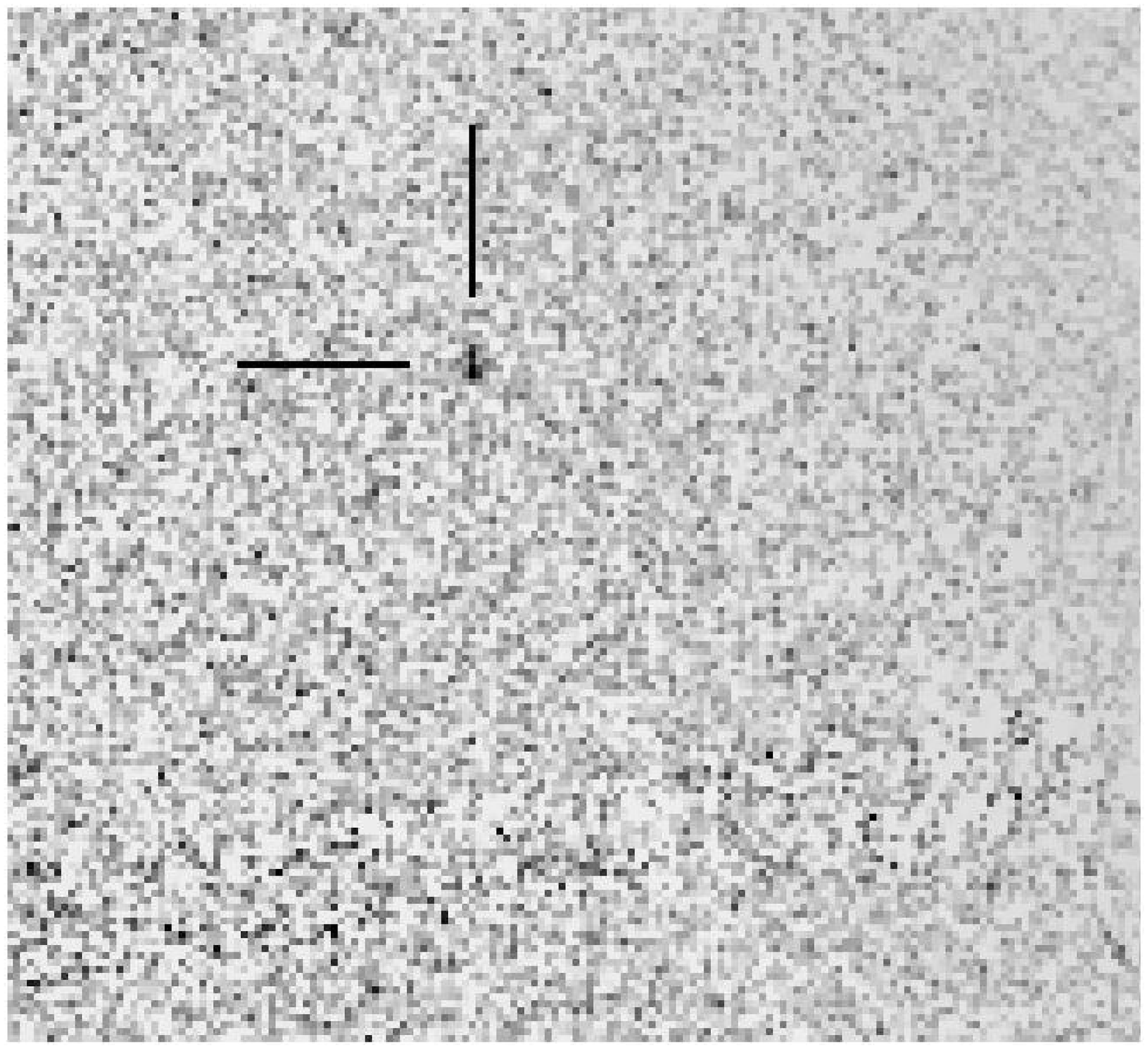}
\caption{The HST/WFPC2 B-band (f450, left) and far-UV (F150LP, right) images of J123652.69+621355.3.  The orientation and alignment is the same in both images and the pointers are in the same location on the sky.  The pointers are each 1$''$ in length and point to the faint source to the North of the target galaxy.  The source is clearly detected in the far-UV image. \label{fig:360}}
\end{figure}

This is the brightest source ($B = 22.41$) in our sample and is undetected in the far-UV.  However, there is a clear ($7\sigma$) detection of a compact source $\sim0.3''$ to the North (Figure \ref{fig:360}).  This smaller object is part of the larger isophote used in computing the optical/near-IR fluxes but is not in our smaller aperture used for far-UV photometry.  It has a flat spectrum (in $f_{\nu}$) across the UV/optical and a far-UV flux $f_{700} = 0.030$ $\mu$Jy, with no break in the SED ($FUV-U$ (AB)= 0.07).  Therefore, it is unlikely to be at the same redshift since there is no indication of IGM absorption.  However, it is interesting to note that, if this system was at $z \sim 3$ and was observed with ground-based (ie. low spatial resolution) optical spectroscopy, this object would appear to have escaping Lyman continuum, with $f_{esc,rel} \sim 0.20$.  Though this is the only object in our sample which appears to have foreground contamination, the contamination at higher redshifts will be much higher as the foreground path is larger.  Therefore, it may be important to obtain high-resolution follow-up of high-redshift escape fraction detections to ensure that the flux is not originating from low-luminosity foreground objects.

\subsubsection{J033239.09-274601.8}
There is a bright ($B = 21.4$) QSO at $z=1.22$ \citep[KX4,][]{croom01} which was not included in this analysis but is detected at 4$\sigma$ in the far-UV with $f_{700} = 0.015 \pm 0.004$ $\mu$Jy.  The optical/near-IR photometry is best fit with power law slope $\alpha_{\nu} = -0.85$ (where $f_{\nu} \propto \nu^{\alpha}$), consistent with the average near-UV spectral slope of QSOs \citep{telfer02}.  If we extrapolate this power-law and apply a correction for IGM absorption, we expect to measure a far-UV flux $f_{700} \sim 0.95$ $\mu$Jy.  Therefore, we estimate an absolute escape fraction $f_{esc} \sim 0.02$, a very small value for an unobscurred QSO.  It is also possible, however, that the escape fraction is high, but that a high column density absorber lies along the line of sight to the QSO, as is expected for $\sim 20$\% of the LOSs at this redshift (see Figure \ref{fig:igm_hist}).  

\section{Discussion}
\label{discussion}

\subsection{Comparison of SED Fits to Flux Ratios}
\label{comp_sed}
The best fit ages and color excesses span the allowed ranges, with median values of $t = 300$ Myr and $E(B-V)=0.19$, similar to the values found in LBGs \citep{shapley01,papovich01}.  We used the best fit SEDs to compute the $(f_{1500}/f_{700})_{stel}$ flux ratio and compare with the assumed value of 8 used in section \ref{ratios}.  The intrinsic flux ratio spans the range of 6.1-11.2 with a median of 9.6, 20\% higher than our assumed average value but within the range of uncertainty of other parameters.  We conclude that simply correcting the measured flux ratio with an average $(f_{1500}/f_{700})_{stel} \sim 8-10$ is not unreasonable for galaxies with ongoing star-formation (eg. LBGs).  The distribution of relative escape fractions using both methods (constant UV-to-LC ratio and UV-to-LC ratio from the SED fits) are plotted in Figure \ref{fig:fesc_hist}, and show very similar distributions.  In the subsequent analysis of individual objects, we use the values derived from the SED fits.  

We have computed the UV-to-LC ratio at 900\AA, $(f_{1500}/f_{900})_{stel}$, from our fits to compare with the spectroscopic studies, finding a median value of 7.3.  This is more than a factor of two larger than the value of $\sim3$ assumed by \citet{steidel01} and \citet{shapley06} and would imply an average relative escape fraction $\sim2$ times higher than those estimates.  This correction to the \citet{steidel01} measurement would yield an unphysical relative escape fraction greater than unity.  Their sample is from the bluest quartile of the total LBG population and may be amongst the youngest, thus decreasing the expected intrinsic break (See Figure \ref{fig:age_ratio}).  However, it is difficult to imagine how the incidence of star-formation can be synchronized across entire galaxies (several kpcs) such that the the total star-formation in the galaxy is less than 10 Myr old.  The \citet{shapley06} sample is more indicative of the total LBG population, where larger values for the intrinsic Lyman break should be used when computing the average $f_{esc,rel}$.  However, our fits show a factor $\sim2$ variation in the break amplitude, demonstrating the need to deduce the starburst age when analyzing individual galaxies.

\subsection{HI Column Density Limits}
A fundamental parameter in escape fraction studies is the relative importance of dust and HI in attenuating the Lyman continuum.  If we assume a Calzetti extinction curve and extrapolate to from 1200\AA\ to 700 \AA\ where the extinciton curve is not empirically determined, we can use our best fit values for dust extinction to determine the level of attenuation at 700 \AA.  If our measured flux limit is still lower than the expected flux after dust attenuation then there must be additional absorption by neutral Hyrdrogen.  We can then derive lower limits to the HI column density within the ISM of these galaxies.  These HI ISM transmission limits are given as exp$(-\tau_{HI,ISM})$ in Table \ref{tab:results}.  Our large lower limits to the observed $f_{1500}/f_{700}$ ratio (after correcting for the intrinsic break and IGM opacity) can be explained entirely by differential dust extinction in all but four of the galaxies.  The limits to the HI transmission for these four galaxies are 0.45, 0.61, 0.78, 0.88 and are not strong limits to the HI column densities.  Significantly deeper observations are needed to determine whether dust or HI is the principal cause of Lyman continuum absorption within the galaxies' ISM.

\subsection{Foreground Contamination}
One of our galaxies (J123652.69+621355.3) has a far-UV detected object within 0.3$''$ of the aperture.  These two objects would appear as one if observed with lower spatial resolution.  Therefore, it is possible to have a foreground object contaminate the photometry (or spectroscopy) of a high redshift galaxy so that it appears to be emitting in the Lyman continuum.  This is especially true with ground-based studies where the spatial resolution is low.  The only detections of escaping Lyman continuum have been found at $z \sim 3$ with ground-based studies so care must be taken to ensure that foreground contamination is not a problem.

Galaxies at $z \sim 1$ have larger angular diameters than galaxies at $z \sim 3$, which increases the likelihood of contamination.  However, the comoving line-of-sight distance to $z \sim 3$ is more than twice the line-of-sight distance to $z \sim 1$ and the space density of UV-luminous galaxies is larger between $1<z<3$ than it is at lower redshift \citep{arnouts05}.  Therefore some contamination in $z \sim 3$ studies might be expected.

We can use galaxy number counts in the $U$-band to determine the likelihood of foreground contamination of $z \sim 3$ as these galaxies must reside in the foreground, $0 < z < 3$, and are emitting at wavelengths that mimic Lyman continuum of LBGs.  Because the expected $(f_{1500}/f_{900})_{obs}$ is so high, even very faint foreground sources ($U \sim 28$) can cause $L^*$ LBGs to appear to have large escape fractions.  The surface density of objects with $U(AB) < 28$ is $\sim 3.5 \times 10^5$ deg$^{-2}$ (Dolch et al 2007, in prep).  If we assume that ground-based imaging or spectroscopy can not resolve objects which lie within a 0.5$''$ radius of each other, than we would expect that each $z\sim3$ galaxy has a $\sim 2$\% chance of foreground contamination.  Given this probability, there is a $\sim22$\% chance that one galaxy (out of 14) in the \citet{shapley06} study is subjected to foreground contamination, but only a $\sim 3$\% chance that the both detections are contaminated.  Therefore, it is unlikely that foreground contamination can entirely explain the large escape fractions at $z \sim 3$.  However, it is important to keep in mind that only a small percentage (maybe $\sim 10-30$\%) of LBGs appear to exhibit large escape fractions.  Although foreground contamination will only cause a few percent of all LBGs to appear to have escaping Lyman continuum, they may comprise a significant percentage of LBGs chosen to have high escape fractions ($\sim 10-20$\%).  This demonstrates the need for high resolution follow-up imaging of sources to confirm that there is no contamination.  

\subsection{Combined Analysis with Other $z \sim 1$ Studies}
\label{z1_comp}
\citet[herafter M03]{malkan03} conducted a similar study to ours, observing 11 luminous starbursts at $1.1<z<1.5$ with STIS and obtained no detections.  They use a significantly broader and bluer filter (F25SRF2) with a $\lambda_{eff} = 1453$\AA, $\sim$ 150 \AA\ shorter than our observations.  Accounting for the mean redshifts of the surveys, this corresponds to $\lambda_{rest} \sim 660$ \AA\  for M03 and $\lambda_{rest} \sim 715$ \AA\ for our study.  Therefore, the flux ratios of M03 are subject to larger effects from dust attenuation and star formation history than those measured here.  Although the STIS camera is less sensitive than the ACS/SBC, the broader filter increases the sensitivity so that the M03 $f_{700}$ limits are close to our derived limits.  In addition, they targeted more luminous starbursts, resulting in limits to the $f_{esc,rel}$ that are $\sim2$ times lower than ours.  The limits in Table 2 of M03 are $1\sigma$ limits and are an additional factor of two too low, as the radius of the aperture used for the limits was smaller, by a factor of two, than the intended 0.5$''$ aperture radius (M. Malkan, private communication).   After applying corrections for IGM absorption (which M03 did not do), we derive limits of $0.10<f_{esc,rel}(3\sigma)<0.42$.  

We have fit the M03 galaxies to SED models and verified that the constant star-formation assumption of $(f_{1500}/f_{700})_{stel}$ is appropriate for their sample.  The STIS field-of-view is small so there are no objects (target galaxy or otherwise) detected in 9 of 11 pointings.  The far-UV images can not be astrometrically aligned with optical data, leaving their measurements subject to the intrinsic pointing uncertainties of HST.   Because these pointing uncertainties are larger than their aperture radii ($1''$ vs. $0.5''$), the stacked far-UV images of the 11 galaxies spread the light over larger areas and do not give significantly better limits than the individual images. 
 
Combining with the M03 sample, we now have 28 galaxies with $f_{esc,rel} \leq 1.0$ and 20 galaxies with $f_{esc,rel}<0.5$ at $z \sim 1$.  The M03 sample have UV luminosisites similar to L$^*$ LBGs at $z\sim3$, whereas our sample is somewhat fainter (0.1-1.0 L$^*_{LBG}$).  These galaxies span nearly two orders of magnitude in luminosity and have a broad range in morphologies and starburst ages, yet we see no evidence for large escape fractions at this redshift.

\begin{figure}
\epsscale{1.0}
\plotone{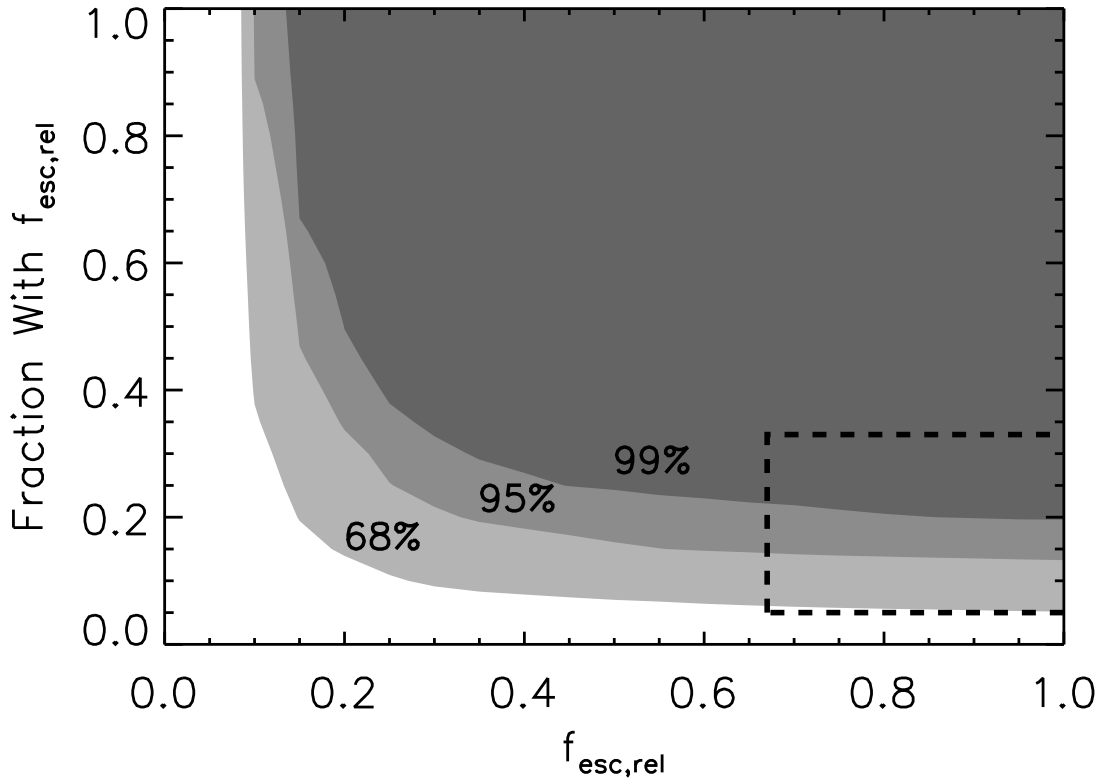}
\caption{The parameter space excluded by a Monte-Carlo analysis of the combined limits of this work with \citet{malkan03}.  The x-axis is the relative escape fraction and the y-axis is the fraction of galaxies which have this escape fraction.  The other galaxies are assumed to have negligible escape fractions.  The shaded regions are excluded(at 3, 2, 1$\sigma$ darker to lighter) by the fact that we detect no galaxies in the combined sample of 32 galaxies.  The dashed box denotes the approximate parameter space implied by \citet{shapley06}.  The horizontal errors to the box are from uncertainties in the determination of the relative escape fraction and the vertical errors are Poisson. \label{fig:mc}}
\end{figure}

As shown in Section \ref{igm}, the transmission through the IGM can vary substantially along different lines-of-sight.  This will affect our results such that $\sim20$\% of our limits are effectively meaningless as they are looking at opaque lines of sight.  However, $\sim 60$\% of our limits will be substantially better than the limits given in Table \ref{tab:results} as these lines-of-sight are more transparent than average.  To better evaluate the effects of this distribution on our survey, we've performed a Monte-Carlo simulation where we observe 32 galaxies (21 from this sample and 11 from M03) with the same distribution of $(f_{1500}/f_{700})_{obs}$ limits as in our surveys.  The IGM transmission is chosen randomly from the distribution plotted in Figure \ref{fig:igm_hist}.  We assume an intrinsic break of $(f_{1500}/f_{700})_{stel}=8$ for all galaxies.  We then need to assume an escape fraction for each galaxy.  For this we assume that a fraction Y of star-forming galaxies at our redshifts have an $f_{esc,rel} = X$, and all other galaxies have effectively no Lyman continuum transmission.  This analysis assumes that this parameter space (X,Y) is the same for galaxies with different luminosities.  We allow X and Y to vary between 0.0 and 1.0 and run 10,000 iterations of our observations for each parameter set (X,Y).  We then determine, for each parameter set, what percent of the time we achieve a null result for all 32 galaxies.  The results are plotted in Figure \ref{fig:mc}.  The shaded regions denote the parameter space which is excluded by our combined samples at 68, 95, and 99\% confidence (from lighter to darker).  If all star-forming galaxies at $z \sim 1.3$ have the same relative escape fraction, Figure \ref{fig:mc} shows that it must be less than $f_{esc,rel} < 0.14$ at 99\% confidence.  Conversely, if some galaxies have large relative escape fractions, $f_{esc,rel} \ge 0.75$, while the others have none, they must be less than 20\% of the total population (at 99\% confidence).  

\subsection{Uniform Comparison of Previous Studies}
In Table \ref{tab:compare}, we summarize the escape fraction studies at all redshifts by converting the results to common definitions: the UV-to-LC ratio corrected for IGM absorption, $(f_{1500}/f_{900})_{corr}$, and the relative escape fraction.  

Because observations at $\lambda_{obs} \sim 1000$\AA\ are so difficult, the local sample consists of only six objects.  Once the Milky Way HI \citep{hurwitz97} and foreground dust extinction \citep[see corrections in][]{deharveng01} are properly accounted for, and the measurements are converted to relative escape fractions, the data do not put strong limits on the relative escape fraction.  In addition, three of the four galaxies in the \citet{leitherer95} sample have very significant color excess, $E(B-V) > 0.5$.  If the Lyman continuum is subject to the same dust extinction, then it would not be detectable in these measurements.  The most sensitive limits at low redshift come from FUSE spectra of Mrk 54 \citep{deharveng01} and Haro 11 (Grimes et al. 2007, submitted) which give UV-to-LC limits of 112 and 21 respectively.  Note that \citet{bergvall06} originally reported a strong Lyman continuum detection, but a recent reanalysis by Grimes et al. (2007, submitted) detect no flux blueward of the Lyman limit.  

The $z\sim 0$ data set is too small to rule out large escape fractions in a small subset of galaxies, and the limits are too weak to rule out small escape fractions in a high percentage of galaxies.  A larger, deeper study of low redshift starbursts (perhaps the UV-luminous galaxies of \citet{hoopes06}) is required for better comparison with the high redshift studies.  

The $z\sim3$ studies claim markedly different conclusions and warrant a more detailed discussion.  In Table \ref{tab:compare}, we have converted the results of the studies to common quantities.  

\citet{steidel01} found a significant (nearly $5\sigma$) detection of the Lyman continuum at $\lambda_{rest} = 880-910$ \AA\ in a composite spectrum of 29 $z\sim3.4$ LBGs.  Once correcting for IGM absorption, the implied UV-to-LC ratio, $(f_{1500}/f_{900})_{corr} = 4.5$ implies a relative escape fraction near unity.  Note that the sensitivity could not place significant constraints on any individual LBG.  \citet{steidel01} point out that, because the LBGs were required to be at the high redshift end of the $U$-dropout sample, these 29 LBGs were among the bluest quartile of the entire LBG sample.  Therefore, it might be expected that differential reddening is not a major issue, and the intrinsic Lyman break amplitude is somewhat smaller (due to younger ages).  Regardless, as \citet{steidel01} point out, this is a surprising result which must be verified with deeper spectra.

\citet{shapley06} published much deeper individual spectra of 14 LBGs, of which two showed significant flux below the Lyman limit.  These UV-to-LC ratios (2.9, 4.5) also imply relative escape fractions near unity, while the spectra of the remaining 12 give a $3\sigma$ limits from $f_{esc,rel} < 0.25-1.0$, depending on depth (when using a Lyman break amplitude of 6).  These limits still allow for non-zero escape fractions, but do not display the small UV-to-LC ratios exhibited in the two detection or in the stack of \citet{steidel01}.    

Several other groups have published escape fraction limits of LBGs at $z\sim3$, finding no detections.  Once converting the limits to $3 \sigma$ and using a more conservative Lyman break amplitude, $(f_{1500}/f_{900})_{stel} = 6$, the data from \citet{giallongo02} and \citet{inoue05} do not significantly constrain the relative escape fraction (ie. limits of $f_{esc,rel} \geqsim 1$, See Table \ref{tab:compare}). \citet{fernandez-soto03} have analyzed broadband photometry of 27 galaxies between $1.9<z<3.5$ in the Hubble Deep Field-North and report a $3\sigma$ limit of $f_{esc} < 0.039$.  This low limit, derived from such a large sample, appears to contradict the large escape fractions detected by S01 and S06.  However, there are several differences in the analysis which should be accounted for.  Firstly, the \citet{fernandez-soto03} SEDs do not include any intrinsic Lyman Break in the stellar SEDs, so the limits need to be multiplied by the additional $(f_{1500}/f_{700})_{stel} = 8$ (the value listed in Table \ref{tab:compare} accounts for this).  Secondly, the filter used to measure the escaping Lyman continuum contains significant flux from redward of the Lyman break, necessitating very accurate determinations of the UV continuum flux in order to estimate the level of Lyman continuum in the same filter.  Once these effects are accounted for, their stacked limit is above the average detection of S06.  In summary, \citet{shapley06} is the only large study at $z \sim 3$ which significantly constrains the escape fraction, and it confirms the preliminary detection of significant ionizing emissivity at $z\sim3$ by \citet{steidel01}.  

The best limits on the relative escape fraction at $z \sim 1$ (Section \ref{z1_comp}) are significantly lower than the average {\it detections} at $z\sim3$ \citep{steidel01,shapley06}, and yet we find no detections in the present survey.  In Figure \ref{fig:mc}, we show the parameter space implied by the S06 results (dashed box) does not agree, at $1\sigma$, with our results, implying that there may be an evolution in the relative escape fraction with redshift.  However, many of the galaxies in our combined sample are somewhat less luminous than the $z\sim3$ studies ($L_{UV}=0.1-1.0 L^*_{UV}$ for $z\sim3$ LBGs).  It is possible that higher UV luminosities, and thus star-formation rates, can create supernova chimneys \citep{fujita03} which clear the surrounding ISM or that the high Lyman continuum flux can ionize low HI column density sight lines.  Therefore, the fact that our galaxies are less luminous may help explain why we see no examples of unity escape fraction.  The galaxies in the M03 sample have similar luminosities to S06 so one might expect to detect a couple of galaxies with large relative escape fractions.  However, the M03 sample alone is too small to rule out the parameter space of S06.  

Another difference between the $z\sim1$ and $z\sim3$ studies is that our broadband observations sample shorter wavelengths of the Lyman continuum so they are far more sensitive to the star formation history.  Our assumption of constant star formation is a best-case scenario since this minimizes the amplitude of the $(f_{1500}/f_{700})_{stel}$ ratio.  If there has been any significant decrease in star-formation  within $t<10$ Myr, $f_{700}$ will be significantly lower than $f_{900}$, while having little significant effect on the broadband SED redward of the Lyman Break.  Therefore, it is possible that some of our galaxies will have a larger $(f_{1500}/f_{700})_{stel}$, thus weakening our limits on $f_{esc,rel}$.

\section{Summary}
We have examined deep far-ultraviolet (1600\AA) imaging of the HDF-North and HUDF to search for escaping Lyman continuum flux from 21 star-forming galaxies at $1.1<z<1.5$, finding no detections.  We account for all factors needed to properly convert our limits to relative escape fractions, including intervening IGM absorption and the amplitude of the intrinisic Lyman break of the stellar population.  

By fitting SED models to the optical/near-IR photometry, we estimate the star-formation age and dust reddening parameters for each individual galaxy.  We find that, although there is $\sim50$\% scatter in the intrinsic stellar Lyman Break due to the starburst age, it is reasonable to assume a constant $(f_{1500}/f_{700})_{stel} \sim 8$ or $(f_{1500}/f_{900})_{stel} \sim 6$ for galaxies with SEDs similar to Lyman Break Galaxies.  This value is two times larger than the $(f_{1500}/f_{900})_{stel} = 3$ assumed in many previous studies.  Assuming a reasonable extrapolation of the extinction curve below the Lyman limit, we show that the observed flux decrement at 700 \AA\ can be attributed to dust attenuation and does not require large column densities of HI within the ISM of the galaxies.  Deeper observations are required to determine the relative importance of dust and HI to the escape fraction.

We obtain $3\sigma$ limits better than $f_{esc,rel}<1.0$ in 18 galaxies and $f_{esc,rel}<0.5$ in 9 galaxies, with some limits down to $f_{esc,rel}<0.10$.  Our stacked fluxes give a combined limit of $f_{esc,rel}<0.08$, similar to the sensitivity achieved by \citet[corrected with common assumptions with this study]{malkan03} for more luminous starbursts at the same redshift.  This is the first study to achieve these sensitivities on high redshift starbursts which are less luminous than typical $L^*_{LBG}$.  These stacks give the deepest escape fraction limits achieved at any redshift and demonstrate a paucity of ionizing emissivity in most starbursts at $z \sim 1.3$.  Including the sample of \citet{malkan03}, we now have 28 galaxies at $z \sim 1.3$ with $f_{esc,rel}<1$ and 20 galaxies with $f_{esc,rel}<0.5$ and no detections.  When properly accounting for the broad distribution of IGM transmission with Monte Carlo simulations, we conclude that, at 99\% confidence, less than 20\% of star-forming galaxies at $z \sim 1$ have relative escape fractions near unity.  Conversely, if all of the galaxies have similar relative escape fractions then, at 99\% confidence, they must have $f_{esc,rel} < 0.14$.

When all of the existing $z \sim 3$ studies are uniformly compared, only the \citet{shapley06} study gives significant limits to the relative escape fraction and finds large ionizing emissivity in at least some ($\sim 10-30$\%) LBGs.  The $z\sim1$ studies are sensitive to lower escape fractions than those at $z\sim3$ (even those with detections), and yet no galaxy at $z\sim 1$ is seen emitting significant ionizing flux.  This disagreement is marginally significant ($1 \sigma$), suggesting a possible decrease in the escape fraction with redshift.  We cannot yet rule out the possibility that the discrepancy is the result of differeint observing methods which probe different regions of the Lyman continuum (700 \AA\ vs. 900 \AA).

Further investigation of the evolution of the escape fraction requires larger samples of $z \sim 1$ galaxies that are better analogs of the $z\sim 3$ LBGs.  We have an ongoing program of HST FUV imaging and spectroscopy to obtain better limits on the escape fraction at $z\sim1$ in the GOODS and COSMOS fields.  Other projects, such as stacking of GALEX sources, will also greatly improve the current limits.

\acknowledgments

The authors would like to thank M. Malkan for reanalyzing the results of \citet{malkan03} to facilitate direct comparison with our data.

The research described in this paper was carried out, in part, by the Jet Propulsion Laboratory, California Institute of Technology, and was sponsored by the National Aeronautics and Space Administration (NASA).  Support for programs 9478 and 10403 was provided by NASA through grants from the Space Telescope Science Institute, which is operated by the Association of Universities for Research in Astronomy, Inc., under NASA contract NAS 5-26555.

{\it Facilities:} \facility{Hubble(ACS)}

\bibliography{apj-jour,all_ref}

\clearpage

\pagestyle{empty}

\begin{deluxetable}{lccccccccccccc}
\tabletypesize{\scriptsize}
\rotate
\tablecaption{Far-UV limits and best-fit SED parameters. \label{tab:results}}
\tablewidth{0pt}
\tablehead{\colhead{Name} & \colhead{z$_{spec}$} & \colhead{fuv\tablenotemark{a}} & \colhead{Age} & \colhead{E(B-V)} & \colhead{$f_{1500}$\tablenotemark{b}} & \colhead{$(f_{1500}/f_{700})_{int}$\tablenotemark{b}} & \colhead{$(f_{1500}/f_{900})_{int}$\tablenotemark{b}} & \colhead{$(f_{1500}/f_{700})_{obs}$\tablenotemark{a}} & \colhead{exp(-$\tau _{IGM}$)} & \colhead{exp(-$\tau _{HI}$)} & \colhead{$f_{esc,rel}$} \\
\colhead{} & \colhead{} & \colhead{[$\mu$Jy]} & \colhead{[Gyr]} & \colhead{} & \colhead{[$\mu$Jy]} & \colhead{} & \colhead{} & \colhead{} & \colhead{} & \colhead{} }
\startdata
  J033233.46-274712.4 &  1.298 &  0.007 &  0.300 &   0.44 &   0.10 &   9.94 &   7.29 &  14.76 &   0.51 &  26.73 &   1.32 \\
  J033234.66-274728.0 &  1.438 &  0.007 &  0.300 &   0.10 &   0.69 &  11.21 &   7.29 &  93.06 &   0.41 &   0.61 &   0.29 \\
  J033234.83-274722.1 &  1.316 &  0.017 &  0.300 &   0.25 &   0.61 &  10.02 &   7.29 &  35.54 &   0.50 &   3.04 &   0.57 \\
  J033235.80-274734.8 &  1.223 &  0.011 &  3.000 &   0.44 &   0.09 &   9.47 &   7.35 &   8.67 &   0.55 &  34.87 &   1.97 \\
  J033236.56-274640.6 &  1.414 &  0.007 &  0.300 &   0.15 &   0.26 &  10.80 &   7.29 &  36.35 &   0.43 &   1.98 &   0.70 \\
  J033236.90-274726.2 &  1.318 &  0.017 &  1.000 &   0.20 &   0.53 &  10.10 &   7.35 &  31.13 &   0.49 &   2.50 &   0.66 \\
  J033237.07-274617.3 &  1.273 &  0.023 &  0.030 &   0.35 &   0.43 &   7.62 &   5.64 &  18.47 &   0.52 &   7.90 &   0.79 \\
  J033237.73-274642.7 &  1.307 &  0.009 &  0.300 &   0.15 &   0.42 &   9.98 &   7.29 &  44.85 &   0.50 &   1.21 &   0.44 \\
  J033238.24-274630.1 &  1.216 &  0.008 &  0.300 &   0.40 &   0.14 &   9.56 &   7.29 &  17.32 &   0.56 &  12.69 &   0.99 \\
  J033239.92-274606.9 &  1.295 &  0.027 &  0.300 &   0.35 &   0.56 &   9.94 &   7.29 &  20.39 &   0.51 &   9.90 &   0.95 \\
  J033240.93-274823.6 &  1.244 &  0.008 &  0.300 &   0.15 &   0.42 &   9.61 &   7.29 &  54.25 &   0.54 &   0.88 &   0.33 \\
  J033241.32-274821.1 &  1.318 &  0.042 &  0.300 &   0.15 &   0.85 &  10.03 &   7.29 &  20.28 &   0.49 &   2.72 &   1.00 \\
  J033244.16-274729.5 &  1.220 &  0.010 &  3.000 &   0.25 &   0.28 &   9.65 &   7.35 &  29.60 &   0.56 &   2.91 &   0.59 \\
  J123643.41+621151.6 &  1.241 &  0.012 &  0.300 &   0.15 &   1.28 &   9.58 &   7.29 & 106.89 &   0.53 &   0.45 &   0.17 \\
  J123647.18+621342.0 &  1.314 &  0.009 &  0.100 &   0.35 &   0.40 &   9.33 &   6.80 &  43.17 &   0.50 &   4.54 &   0.43 \\
  J123649.44+621316.6 &  1.238 &  0.008 &  0.300 &   0.44 &   0.33 &   9.56 &   7.29 &  42.88 &   0.54 &   7.61 &   0.41 \\
  J123649.95+621225.5 &  1.204 &  0.006 &  1.000 &   0.20 &   0.24 &   9.58 &   7.35 &  39.92 &   0.57 &   1.52 &   0.42 \\
  J123652.69+621355.3 &  1.355 &  0.024 &  0.010 &   0.30 &   3.20 &   6.11 &   4.38 & 132.09 &   0.47 &   0.78 &   0.10 \\
  J123656.13+621329.7 &  1.242 &  0.018 &  1.000 &   0.15 &   0.66 &   9.66 &   7.35 &  37.54 &   0.54 &   1.28 &   0.48 \\
  J123656.60+621252.7 &  1.233 &  0.013 &  1.000 &   0.10 &   0.38 &   9.58 &   7.35 &  28.06 &   0.55 &   1.22 &   0.62 \\
  J123656.73+621252.6 &  1.231 &  0.010 &  3.000 &   0.44 &   0.12 &   9.55 &   7.35 &  12.33 &   0.55 &  26.01 &   1.41 \\
\enddata
\tablenotetext{a}{3$\sigma$ upper limits.}
\tablenotetext{b}{Derived from SED fits to the optical/near-IR data.}
\end{deluxetable}

\begin{deluxetable}{llcc}
\tablecaption{Compilation of flux ratio limits (corrected for average IGM absorption) and relative escape fractions.  The IGM absorption is redshift dependent so, in order to facilitate comparison between surveys at different redshifts, we have accounted for this by multiplying the flux ratios by the average IGM transmission at the corresponding redshift.  Limits are in parentheses and have been converted to $3\sigma$.  The $z \sim 0$ results also account for foreground extinction using values derived in \citet{deharveng01}.  A conversion of $f_{900} = 1.333 \times f_{700}$ was used based on the constant star-formation SEDs.  A $(f_{1500}/f_{900})_{stel} = 6$ or $(f_{1500}/f_{700})_{stel} = 8$ ratio has been assumed to convert $(f_{1500}/f_{900})_{obs}$ or $(f_{1500}/f_{700})_{obs}$ to $f_{esc,rel}$.   The \citet{fernandez-soto03} limits have been multiplied by eight to account for the intrinsic Lyman break of the stellar population.  \label{tab:compare}}
\tablewidth{0pt}
\tablehead{\colhead{Redshift} & \colhead{Sample} & \colhead{($f_{1500}/f_{900})_{corr}$} & \colhead{$f_{esc,rel}$}}
\startdata
$z\sim0$ 	& 	\citet{leitherer95}	& (12, 7.8, 7.6, 7.5) 	& (0.50, 0.77, 0.79, 0.80)\\
			&	\citet{deharveng01} & (112)				& (0.05) \\
			&	\citet{bergvall06} Haro-11 	& 13 		& 0.46 \\
			&   Grimes et al. (2007) Haro-11 reanalysis & (21) 	& (0.29) \\ 
&&&\\
$z\sim1.3$	&	\citet{malkan03} 	& (14-59) 	& (0.10-0.42) \\
			&	This Work			& (8-47)	& (0.10-1.0) \\
			&   This Work stack  	& (59)		& (0.08) \\
&&&\\
$z\sim3$	&	\citet{steidel01} stack	& $<4.6>$ 	&  $\sim1$ \\
			&   \citet{giallongo02} & (6.2,6.0)	&  (0.95, 1.00) \\ 
			&   \citet{fernandez-soto03} stack &  \nodata & (0.32) \\
			&   \citet{inoue05}		& (2.6,4.0) 	&  (2.3,1.4) \\ 
			&   \citet{shapley06} detections   & 2.9,4.5	&  $\sim 1$ \\
			&	\citet{shapley06} 	& $<22>$  	&  0.27 \\
\enddata
\end{deluxetable}

\clearpage
\pagestyle{plaintop}

\pagestyle{plaintop}

\pagestyle{plaintop}

\clearpage

\pagestyle{plaintop}

\clearpage
\pagestyle{plaintop}

\clearpage
\pagestyle{plaintop}

\clearpage
\pagestyle{plaintop}

\pagestyle{plaintop}

\clearpage
\pagestyle{plaintop}

\clearpage
\pagestyle{plaintop}

\clearpage
\pagestyle{plaintop}

\end{document}